\documentclass[12pt,preprint]{aastex}

\def\Lsun{\hbox{\it L$_\odot$}}

\def\Msun{\hbox{\it M$_\odot$}}
\def\Minit{\hbox{\it M$_{\rm initial}$}}

\def\Myr{\hbox{\it Myr}}

\def\kpc{\hbox{\it kpc}}

\def\kms{\hbox{km$\,$s$^{-1}$}}

\def\AK{\hbox{\it $A_{\rm K}$}}

\def\simgr{\mathrel{\hbox{\rlap{\hbox{\lower4pt\hbox{$\sim$}}}\hbox{$>$}}}}

\shorttitle{RSGs}
\shortauthors{FIGER ET AL.}

\begin{document}

\title{DISCOVERY OF AN EXTRAORDINARILY MASSIVE CLUSTER OF RED SUPERGIANTS}

\author{
Donald F. Figer\altaffilmark{1}, 
John MacKenty\altaffilmark{2},
Massimo Robberto\altaffilmark{2},
Kester Smith\altaffilmark{2}, \\
Francisco Najarro\altaffilmark{3},
Rolf P. Kudritzki\altaffilmark{4},
Artemio Herrero\altaffilmark{5}
}

\email{figer@stsci.edu}

\altaffiltext{1}{Center for Imaging Science, 54 Lomb Memorial Drive, Rochester Institute of Technology, Rochester NY 14623-5604}
\altaffiltext{2}{Space Telescope Science Institute, 3700 San Martin Drive, Baltimore, MD 21218; figer@stsci.edu }
\altaffiltext{3}{Instituto de Estructura de la Materia, CSIC, Serrano 121, 29006 Madrid, Spain }
\altaffiltext{4}{Institute for Astronomy, University of Hawaii, 2680 Woodlawn Drive, Honolulu, HI 96822}
\altaffiltext{5}{Instituto de Astrofísica de Canarias, 38200 La Laguna, Tenerife, Spain; Departamento de Astrofísica, Universidad de La Laguna, 38071 La Laguna, Tenerife, Spain}

\begin{abstract}
We report the discovery of an extraordinarily massive young cluster of stars in the Galaxy, having an
inferred total initial cluster mass comparable to the most massive young clusters
in the Galaxy. Using {\it IRMOS}, {\it 2MASS}, and {\it Spitzer} observations, we conclude that there are 14 
red supergiants in the cluster, compared with five, in what was previously thought to be the richest Galactic cluster of such stars. 
We infer spectral types from near-infrared spectra that reveal deep CO bandhead absorption that can only be fit by red
supergiants. We identify a gap of $\Delta${\it K}$_s$$\sim$4 magnitudes between the stars and the
bulk of the other stars in the region that can only be fit by models if the brightest
stars in the cluster are red supergiants. We estimate a distance of 5.8~\kpc\ to the cluster by
associating an OH maser with the envelope of one of the stars. We also identify a ``yellow'' supergiant of G6~I type in the cluster. 
Assuming a Salpeter IMF, we infer an initial cluster mass of 20,000 to 40,000~\Msun\ for
cluster ages of 7-12~\Myr. Continuing with these assumptions, we find 80\% of the
intial mass and 99\% of the number of stars remain at the present time. 
We associate the cluster with an x-ray source (detected by {\it ASCA} and {\it Einstein}), a recently discovered very high energy $\gamma$-ray 
source (detected by {\it INTEGRAL} and {\it HESS}), and several non-thermal radio sources, finding 
that these objects are likely related to recent supernovae
in the cluster. In particular, we claim that the cluster has produced at least one recent supernova remnant with
properties similar to the Crab nebula. It is not unlikely to find such a source in this cluster, given
our estimated supernova rate of one per 40,000 to 80,000~{\it yr}. 
\end{abstract}
\keywords{stars: evolution --- stars: supergiants --- infrared: stars}
\section{Introduction}
Massive stars have short lives and terminate their nuclear energy generation
in supernovae near the massive stellar clusters in which they were born. 
The relationship between inital stellar mass and post-supernova end state depends 
on the mass lost during the
pre-supernova phases, i.e.\ main sequence, blue supergiant (BSG), red supergiant (RSG), and
WR phases \citep{heg03}. The mass of a star just before collapse as a supernova is the 
critical discriminant that determines the density of the end state, i.e.\ 
complete disruption, neutron star, black hole, and total implosion to black hole with no supernova.
The hydrogen content determines spectroscopic
features of supernovae. Type~II-P/L supernovae are thought to be produced by
progenitors with hydrogen rich envelopes and Type~Ib/c supernovae by those without
hydrogen \citep{heg03}. The mass of the envelope may also influence the spectroscopic
properties of the supernovae; therefore, the wind generated mass-loss history and/or mass
loss driven by common envelope evolution may be important in this regard. Further, rotation
rate and metallicity may be important variables, as they affect wind driven
mass-loss \citep{hir04}.
\clearpage
\begin{figure*}
\epsscale{1}
\vskip 20 cm
Color image can be found at: http://www.cis.rit.edu/\ensuremath{\sim}dffpci/images/f1.eps
\caption
{\label{fig:color} {\it 2MASS} color composite image.}
\end{figure*}

Evolutionary and core collapse models predict that red 
supergiants may be the immediate predecessors to Type~II-P supernovae \citep{heg03}, 
and there is supporting obervational evidence for this scenario \citep{sma04,van05}; although in one particularly famous 
example (1987A), the type is B3~I \citep{wal89}.
The observational evidence is necessarily quite limited, as red supergiants
are relatively rare in the Galaxy. Indeed, there are only $\sim$200 
known in the Galaxy \citep{hum78,gar92,lev05}, and no more than
five in any single coeval cluster \citep{bea94,car03}. 
This scarcity is understandable as a consequence of the Galactic cluster mass function,
the stellar initial mass function, the presence of insterstellar dust in the Galactic disk, and the shortness of the RSG phase
predicted by stellar evolution models. In particular, the Galactic cluster mass
function \citep{kha05} and stellar initial mass function \citep{sal55,kro02} are both decreasing with
mass and RSGs only evolve from relatively massive stars having
\Minit$\sim$8 to 25~\Msun\ and ages of 6 to 15~\Myr. Clearly, it is important
to identify more RSGs, especially in coeval clusters, to test stellar
evolution, core collapse, and end state predictions. 

Much of our knowledge of RSG populations comes from studies of nearby galaxies and nearby regions in the Galaxy.
\citet{mcg84} and \citet{mas03} discuss RSGs in open clusters in the SMC, LMC, and Galaxy.
\citet{hum70} lists RSGs in Galactic associations. 
\citet{hum84} discuss the BSG to RSG ratio in Galactic clusters.
\citet{hum79a} gives histograms of the number of RSGs versus spectral type in the
SMC, LMC, and Galaxy, showing that RSGs in the former are significantly earlier than in the latter two. 
\citet{hum79b} gives a list of RSGs in the LMC. The highest number of such stars in
any association in the LMC is eight. In the Galaxy, the Per~OB1 association contains $\sim$20 
RSGs, but it is still unclear as to whether they form a group with a common, and
coeval, formation history \citep{sle02}. 
The richest known coeval cluster of RSGs in the Galaxy is NGC~7419, with five \citep{car03}. 

It is interesting to analyze RSGs in coeval clusters because one can then be
sure that observed differences between the stars are not simply related to age or metallicity effects. Of course, it
is necessary to use massive clusters for such purposes, given that only the most
massive clusters can have a significant number of RSGs. As an example,
a cluster must have an initial mass at least as great as 10$^4$~\Msun\ to have
more than 10 RSGs; it must also be at an age when its members
will be in the RSG stage. 
In the Galaxy, there are only a few clusters with masses greater than 
10$^4$~\Msun. They are the Arches cluster \citep{fig02}, the Quintuplet cluster \citep{fig99a,fig99b},
the Central cluster \citep{fig04}, and Westerlund~1 \citep{cla05}. In the former three cases,
the clusters were identified by infrared observations for their close proximity to the Galactic center.
In the latter case, the cluster was identified at optical wavelengths. This leaves a
large portion of the Galaxy that is obscured at optical wavelengths, but has yet to be probed. 

\citet{bic03a,bic03b} and \citet{dut03a} created a catalog of massive cluster candidates using
an algorithm to search the {\it 2MASS} survey \citep{skr97,skr06}\footnote{This publication makes use
of data products from the Two Micron All Sky Survey, which is a joint project of the
University of Massachusetts and the Infrared Processing and Analysis Center/California
Institute of Technology, funded by the National Aeronautics and Space Administration
and the National Science Foundation.}. Most of these candidates are obscured at optical
wavelengths, and some have already been
shown to be massive stellar clusters \citep{dut03b,bic04,bor05,iva05,lei05}.

Using this catalog, we have identified one of the most massive stellar clusters in the
Galaxy. In this paper, we present infrared photometry and spectra, and an analysis using
multiwavelength data sets that strongly suggests
that there is a very massive young stellar cluster containing 14 RSGs near G25.25$-$0.15. 

\section{Observations and Data Reduction}
Spectra were obtained in September and October of 2005 at the {\it KPNO}~2.1m and 4m~telescopes
using {\it IRMOS} \citep{mac04}. The spectral resolution was $\sim$1,000, 
using a 3 pixel wide slit aperture.  
Calibration files (source darks, flat and Neon lamps and relative darks) 
were taken immediately after the sequence of source exposures for the 
same slit configuration. The spectra were extracted from background-subtracted
and flat-fielded images. The wavelength scale was fixed by relating pixel number to
the locations of sky OH and Neon emission lines in the calibration frames. The resultant
S/N was generally greater than 100, as estimated from the data. 

We also extracted {\it 2MASS} images and point source photometry from {\it 2MASS} catalog, 
on a field centered at RA=18$^h$37$^m$58$^s$, DEC=$-$6$\arcdeg$52$\arcmin$53$\farcs$0 (J2000).
We obtained {\it Spitzer/IRAC} photometry through the {\it GLIMPSE} ({\it Galactic Legacy Infrared Mid-Plane Survey Extraordinaire}) 
point source catalog \citep{ben03}\footnote{This work is based in part on observations made 
with the {\it Spitzer} Space Telescope, which is operated by the Jet Propulsion Laboratory, California 
Institute of Technology under a contract with NASA.}.
\clearpage
\begin{deluxetable}{clrrrcccc}
\tablewidth{0pt}
\tablecaption{Supergiant Cluster Stars}
\tablehead{
\colhead{ID} &
\colhead{{\it 2MASS} Name} &
\colhead{J} & 
\colhead{H} & 
\colhead{{\it K}$_s$} &
\colhead{Sp. Type} &
\colhead{Obs. Date} &
\colhead{{\it M}$_Ks$} &
\colhead{{\it Log(L/\Lsun)}} 
}
\startdata
1 & 18375629-0652322 & 9.748 & 6.587 & 4.962 & M3~I & 16-Sep-05 & $-$11.60 & 5.6 \\
2 & 18375528-0652484 & 9.904 & 6.695 & 5.029 & M4~I & 16-Sep-05 & $-$11.53 & 5.6 \\
3 & 18375973-0653494 & 9.954 & 6.921 & 5.333 & M4~I & 16-Sep-05 & $-$11.23 & 5.5 \\
4 & 18375090-0653382 & 9.658 & 6.803 & 5.342 & M0~I & 16-Sep-05 & $-$11.22 & 5.5 \\
5 & 18375550-0652122 & 10.547 & 7.178 & 5.535 & M6~I & 13-Oct-05 & $-$11.03 & 5.4 \\
6 & 18375745-0653253 & 9.866 & 7.038 & 5.613 & M5~I & 13-Oct-05 & $-$10.95 & 5.4 \\
7 & 18375430-0652347 & 9.941 & 7.065 & 5.631 & M2~I & 16-Sep-05 & $-$10.93 & 5.4 \\
8 & 18375519-0652107 & 10.772 & 7.330 & 5.654 & M3~I & 13-Oct-05 & $-$10.91 & 5.4 \\
9 & 18375777-0652222 & 10.262 & 7.240 & 5.670 & M3~I & 16-Sep-05 & $-$10.89 & 5.3 \\
10 & 18375952-0653319 & 10.179 & 7.218 & 5.709 & M5~I & 13-Oct-05 & $-$10.85 & 5.3 \\
11 & 18375172-0651499 & 10.467 & 7.325 & 5.722 & M1~I & 16-Sep-05 & $-$10.84 & 5.3 \\
12 & 18380330-0652451 & 10.143 & 7.238 & 5.864 & M0~I & 16-Sep-05 & $-$10.70 & 5.3 \\
13 & 18375890-0652321 & 10.907 & 7.716 & 5.957 & M3~I & 15-Sep-05 & $-$10.61 & 5.2 \\
14 & 18374764-0653023 & 10.495 & 7.576 & 6.167 & M1~I & 16-Sep-05 & $-$10.40 & 5.4 \\
15 & 18375778-0652320 & 10.651 & 8.070 & 6.682 & G6~I & 15-Sep-05 & $-$9.88 & 5.0 \\
\enddata
\tablecomments{The spectral types, absolute magnitudes, and luminosities are estimated in this paper.}
\end{deluxetable}
\clearpage
\begin{figure*}
\epsscale{1.5}
\plotone{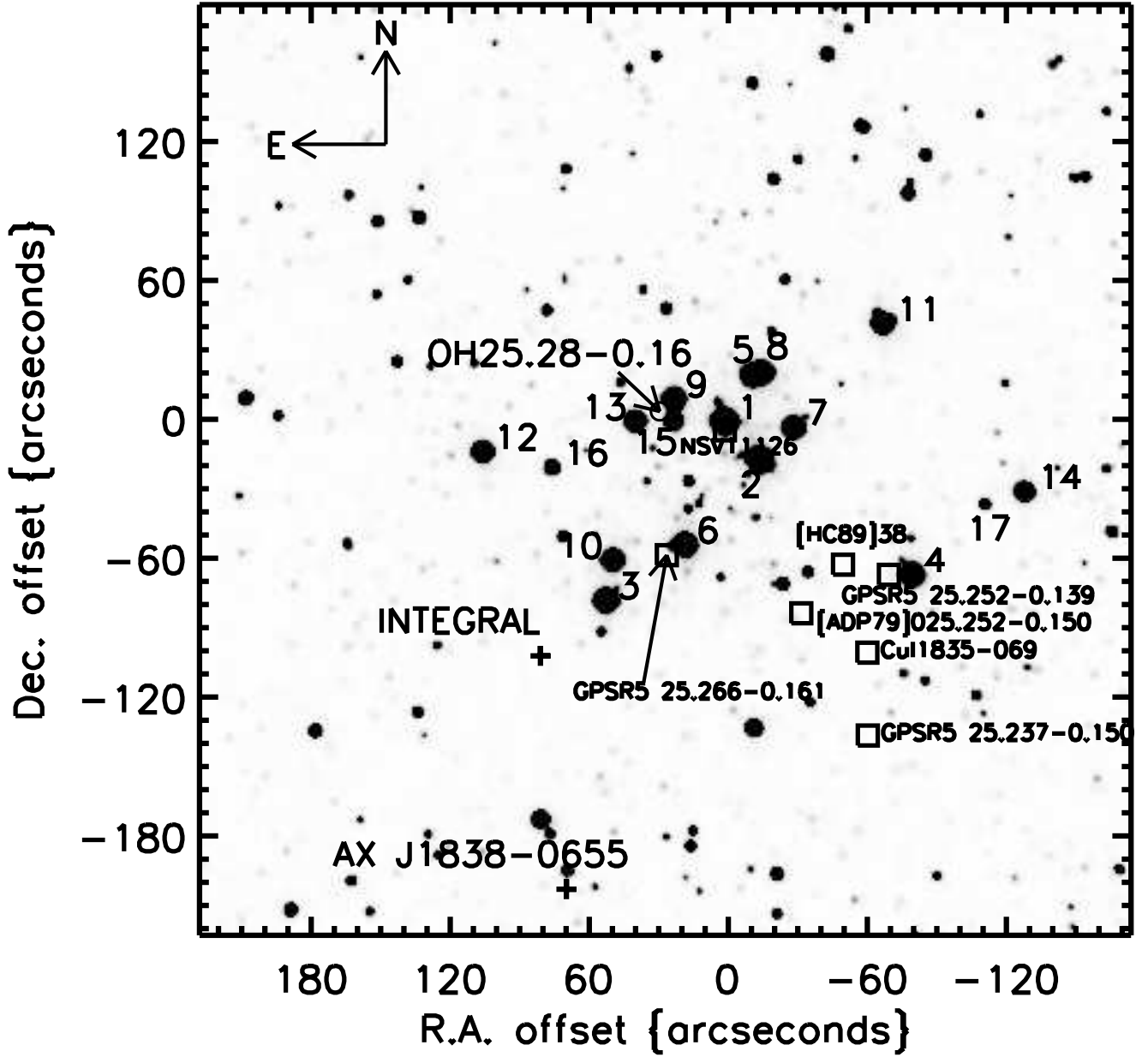}
\caption
{\label{fig:cont} {\it 2MASS} {\it K}$_s$-band image of cluster with offset coordinates from the
brightest star in the cluster (\#1). Stars are identified with numbers according
to their IDs in Table~1. The maser OH~25.28$-$0.16 is designated by
an open circle, and its location has a positional uncertainty of 4$\arcsec$ \citep{blo94}. 
The plus signs indicate the most likely positions of the {\it INTEGRAL}
source and an {\it ASCA} source, AX~J1838.0$-$0655, with positional uncertainties of 3\arcmin\ and 1\arcmin,
respectively. Other objects, taken from {\it SIMBAD}, are designated by squares. Most
of these objects are radio sources.}
\end{figure*}

\begin{figure*}
\epsscale{1.5}
Plot can be found at: http://www.cis.rit.edu/\ensuremath{\sim}dffpci/images/f3.eps
\caption
{\label{fig:msx} {\it MSX} 8.3~\micron\ image of cluster with offset coordinates from the
brightest star in the cluster (\#1). The W42 star formation region is located about 5\arcmin\ to
the northeast of the cluster. The {\it HESS} source has a positional uncertainty of 1-2\arcmin, and 
a size of 7\arcmin.}
\end{figure*}

Table~1 lists the coordinates, magnitudes, spectral types, absolute infrared magnitudes, and luminosities, for the
target stars. The coordinates are taken from the {\it 2MASS} Point Source Catalog, and the other quantities
are determined in the remainder of this paper. 

Fig.~\ref{fig:color} shows a color composite image from the
{\it 2MASS} survey. Fig.~\ref{fig:cont} shows a {\it K}$_s$-band image from the
{\it 2MASS} survey, with designations for the objects in the table and objects extracted
from {\it SIMBAD}\footnote{This research has made use of the {\it SIMBAD} database,
operated at CDS, Strasbourg, France}. Many of these sources from the {\it SIMBAD} database
were observed at x-ray, $\gamma$-ray, and radio wavelengths, as is discussed later in this paper. Fig.~\ref{fig:msx} shows the region around the
cluster at mid-infrared wavelengths, as observed by the {\it Midcourse Space Experiment} {\it MSX} mission \citep{pri03}. 

\begin{figure*}
\epsscale{0.9}
\plotone{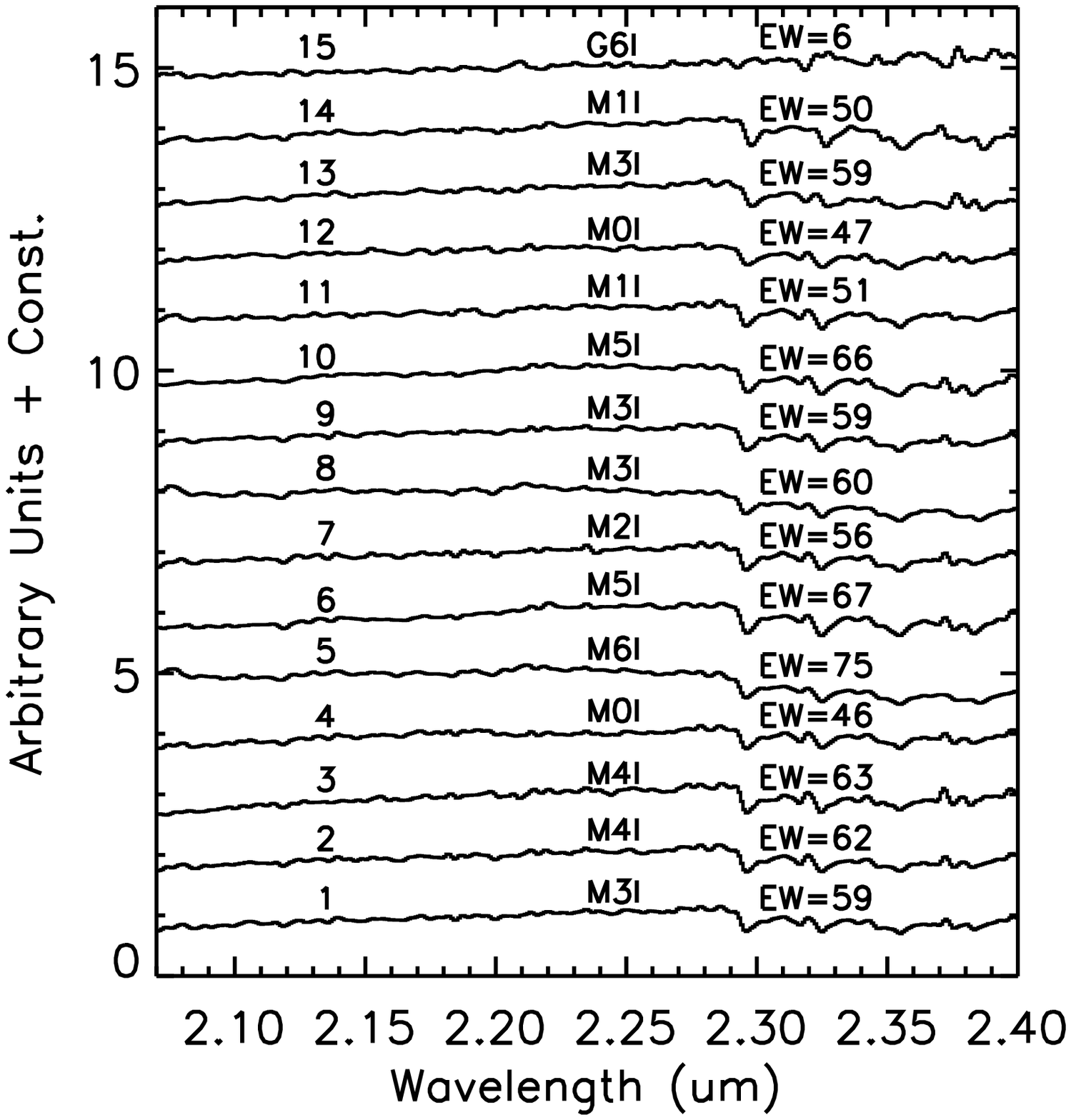}
\caption
{\label{fig:spec} {\it IRMOS} spectra of RSG cluster stars.  The equivalent width
of the $^{12}$CO bandhead, in angstroms, and the corresponding spectral subtype, are given. 
Note that all spectra show deep CO absorption,
except for object \#15, which is the next brightest star after the 14 brightest stars
in the cluster. From its spectrum, and photometric color, we suggest that this star is G-type supergiant.}
\end{figure*}

\begin{figure}
\epsscale{0.9}
\plotone{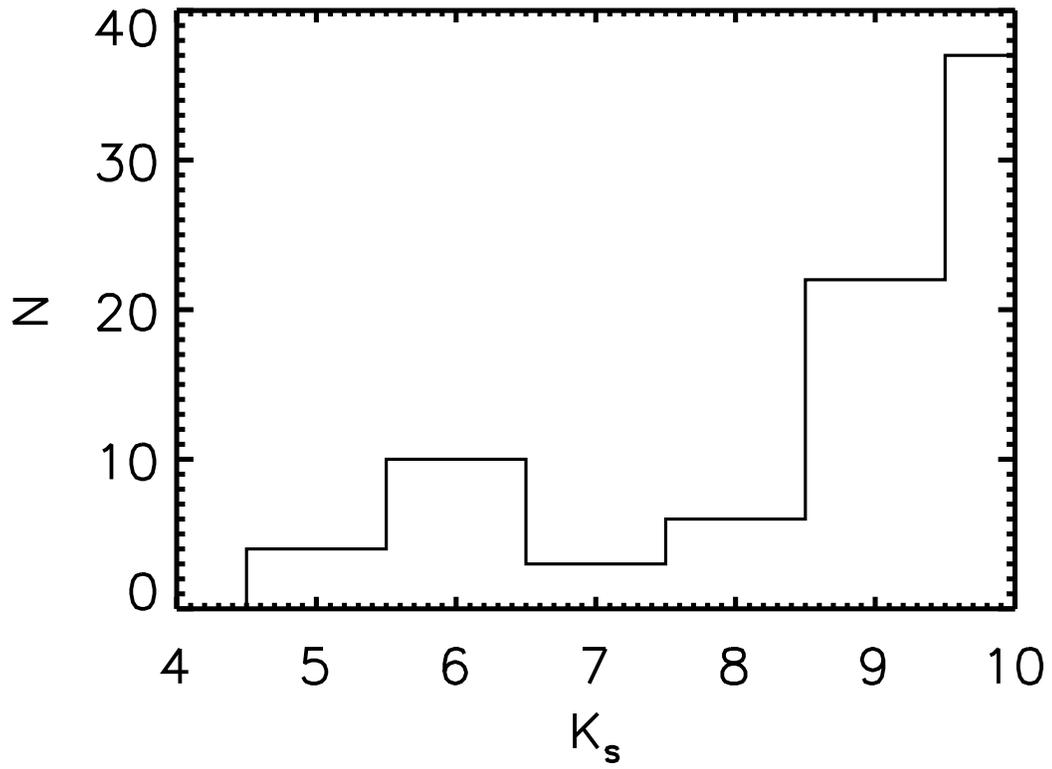}
\caption
{\label{fig:hist} {\it 2MASS} luminosity function of stars within 3\arcmin\ of the center
of the RSG cluster. The 14 RSGs are in the two brightest bins. Note
the relatively large gap in brightness between the RSGs and the bulk of the fainter stars
in the field. The sample has been culled
of stars with reported errors greater than 0.1 magnitudes in {\it K}$_s$.}
\end{figure}

\begin{figure*}
\epsscale{1.6}
\plottwo{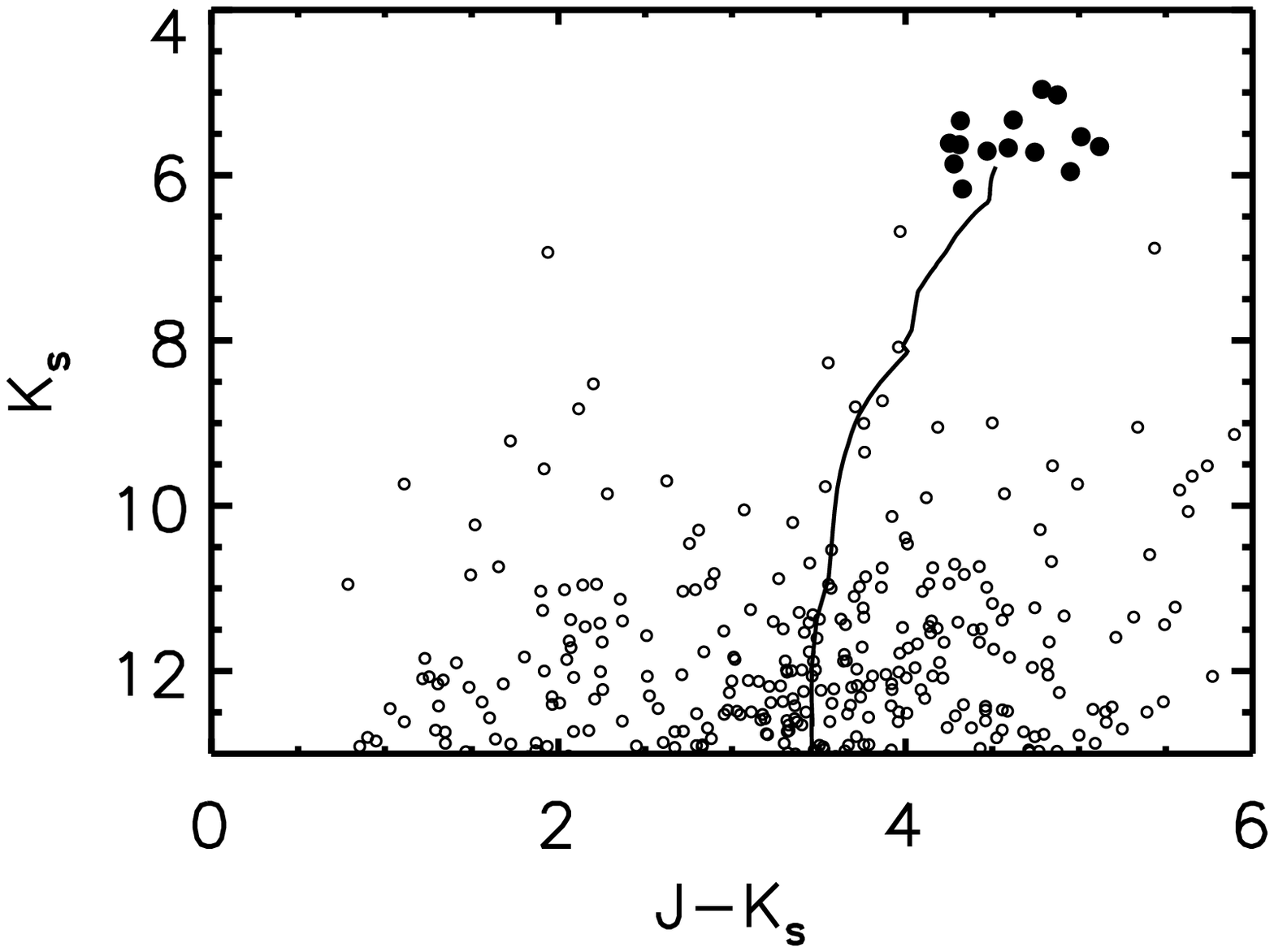}{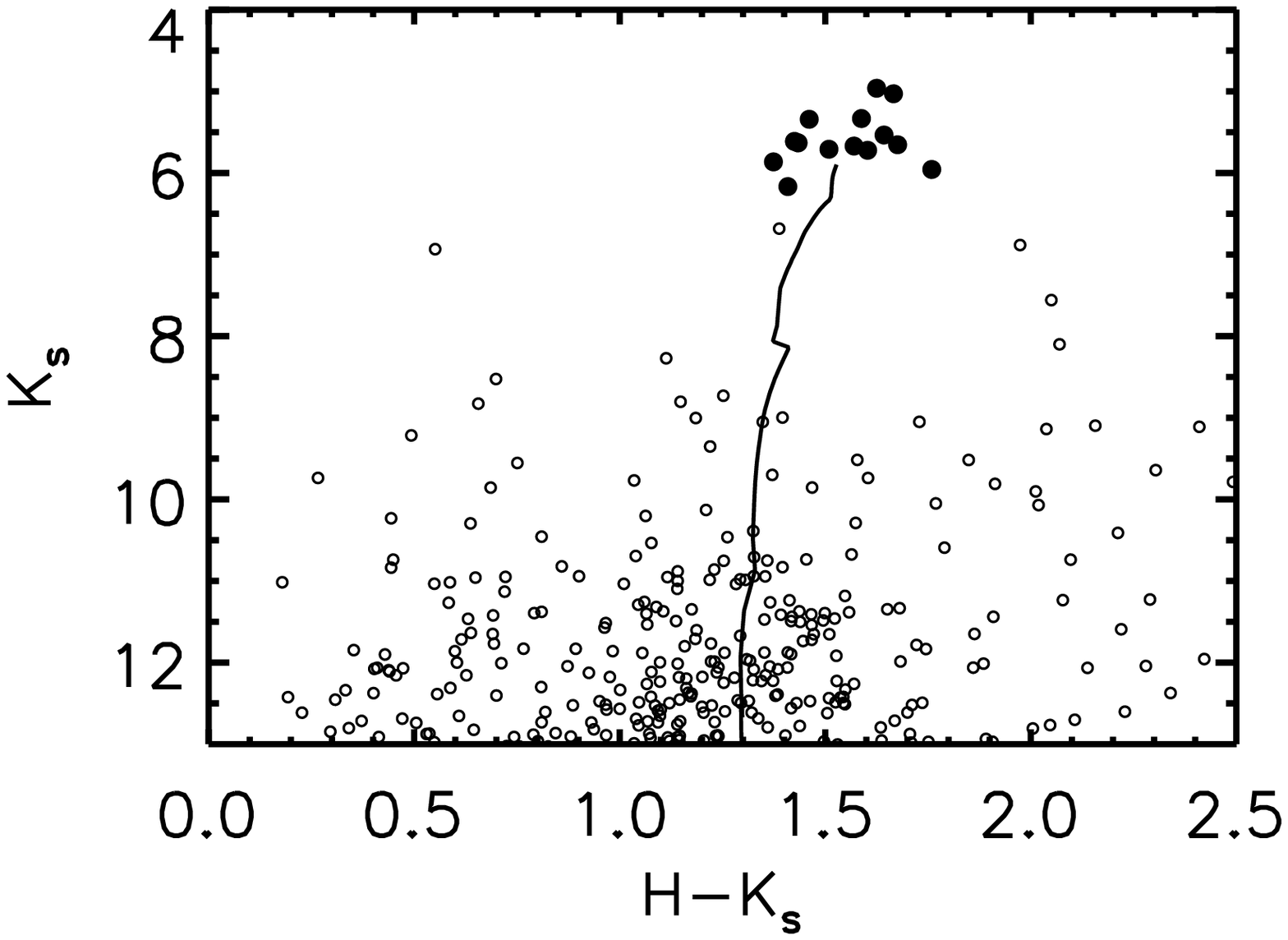}
\caption
{\label{fig:cmd} {\it 2MASS} color-magnitude diagrams of stars within 3\arcmin\ of the center
of the RSG cluster showing the RSGs ({\it filled circles}) and the fainter field stars 
({\it open circles}). A 10~\Myr\ isochrone is overplotted, assuming
the Geneva models with solar metallicity \citep{sch93}, a distance of 5.8~\kpc, \AK$_s$=2.74, and
the redenning law of \citet{rie89}. The color spread for the RSGs scales
roughly with inferred spectral type, i.e.\ later types are redder. The sample has been culled
of stars with reported errors greater than 0.1 magnitudes in {\it K}$_s$.}
\end{figure*}
\clearpage
\section{Analysis}

In this section, we estimate the spectral types, extinction, distance, and luminosities of the cluster stars. 

\subsection{Spectral Types}
The brightest cluster stars appear to be RSGs, having an average spectral type of M3~I, 
consistent with the average type in the Galaxy of M2~I \citep{eli85}. 
Their spectra all have deep CO absorption near 2.2935~\micron, and longer wavelengths, except for star \#15, as shown
in Fig.~\ref{fig:spec}. They are also much brighter than other stars
in the region/cluster, as seen in Figs.~\ref{fig:cont}, \ref{fig:hist}, and \ref{fig:cmd}, and in Table~1.
We measure equivalent widths near the $^{12}$CO bandhead between 2.290~\micron\ and 2.320~\micron, with
the continuum level estimated as the average flux between 2.285~\micron\ and 2.290~\micron. 

\begin{figure*}
\epsscale{1.6}
\plottwo{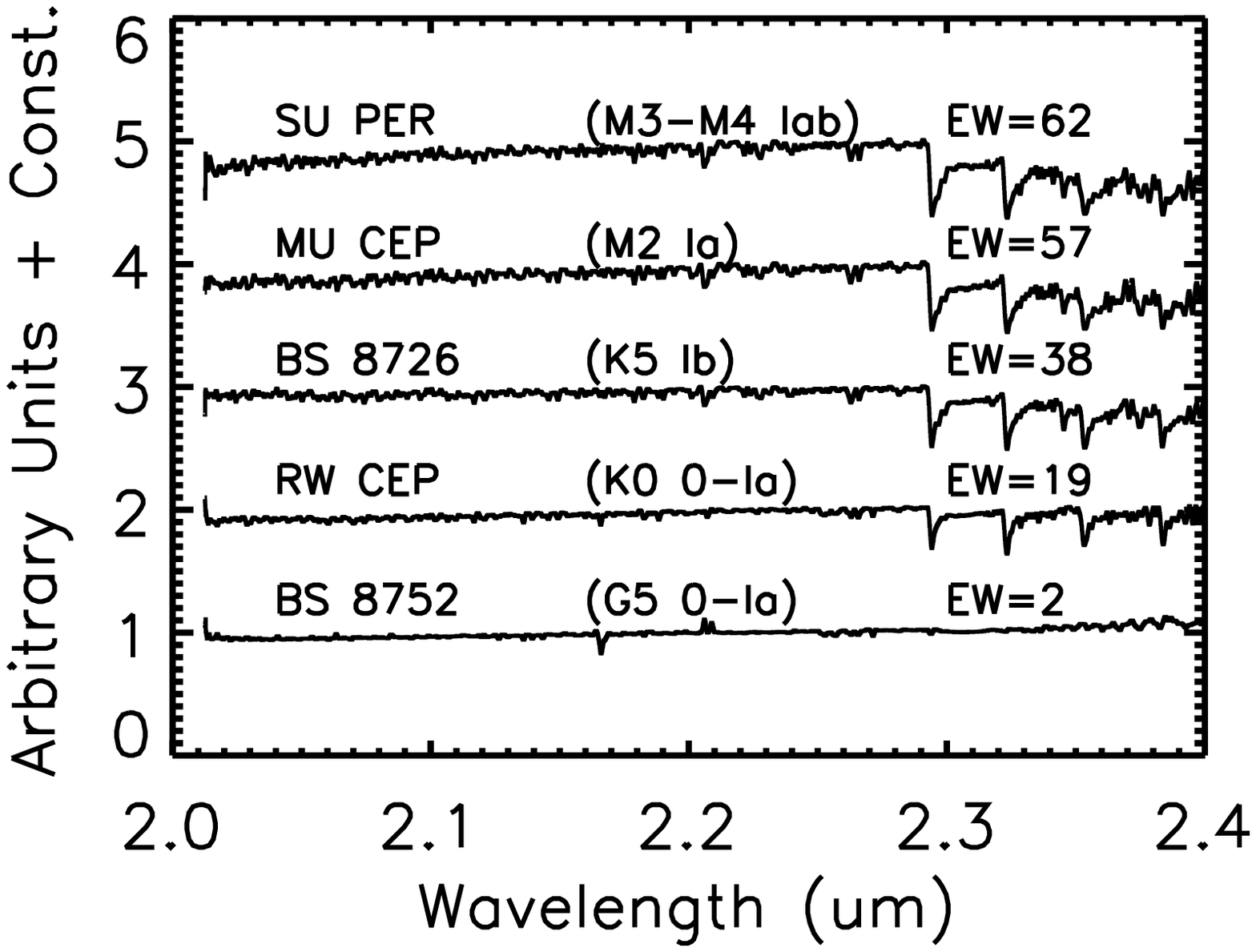}{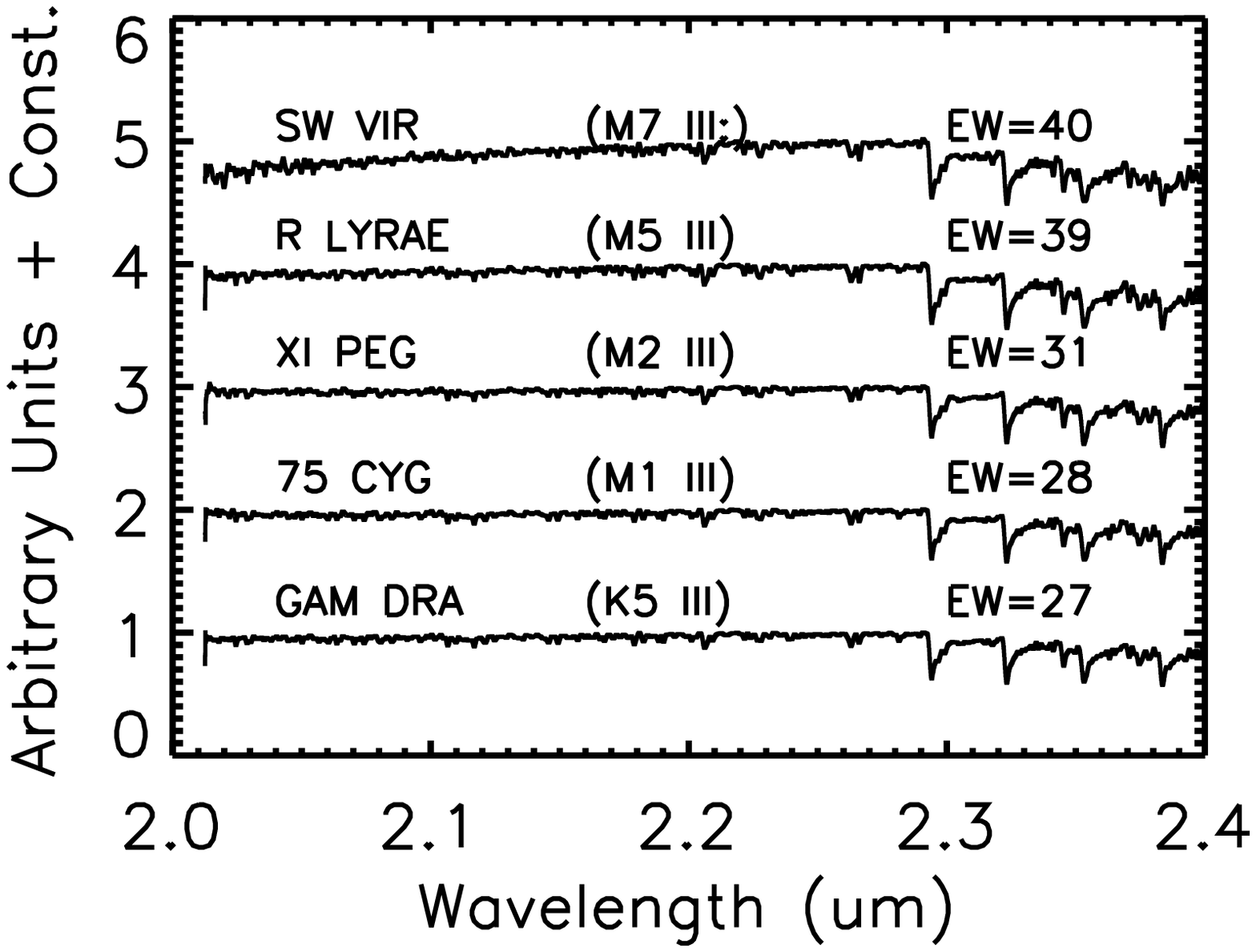}
\caption
{\label{fig:kh} Spectra of template RSGs ({\it left}) and red giants ({\it right}) from \citet{kle86}. 
The equivalent width of the $^{12}$CO bandhead, in angstroms, is given.}
\end{figure*}

\begin{figure}
\epsscale{1.1}
\plotone{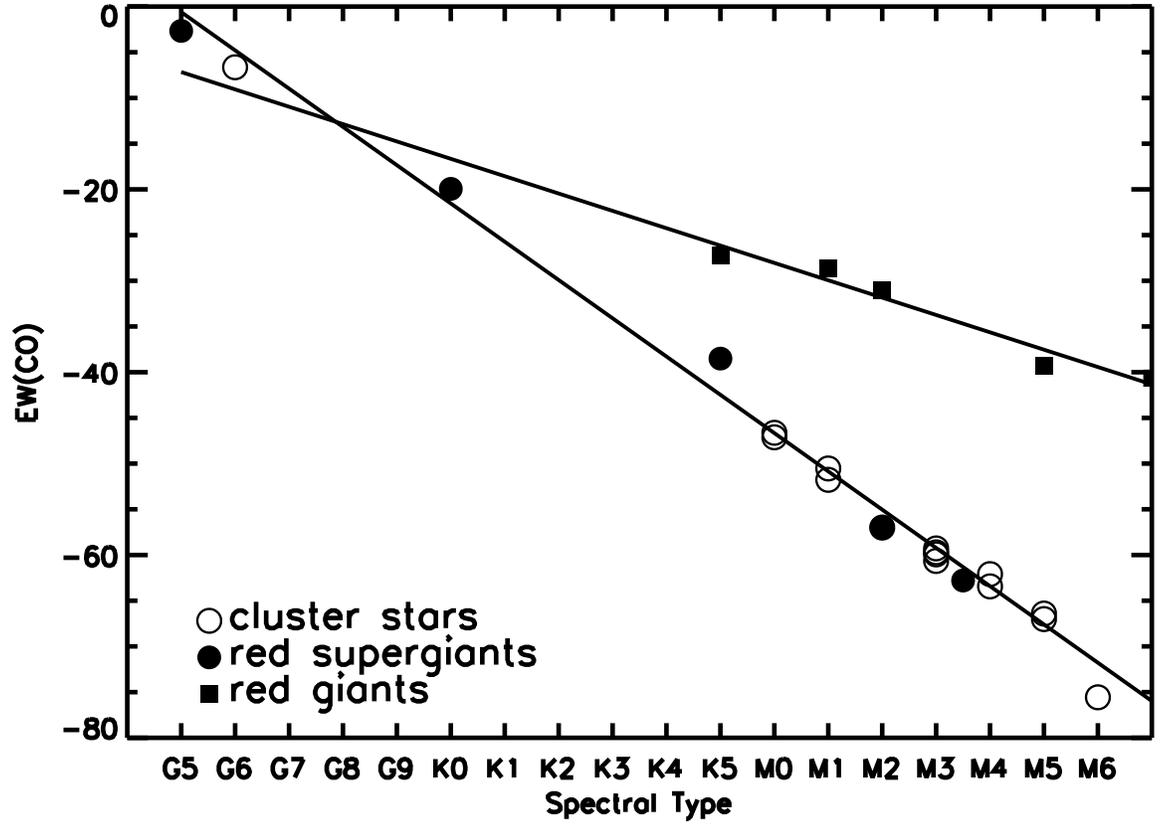}
\caption
{\label{fig:kh_relation} Relation between $^{12}$CO equivalent width, as defined in the text, and spectral subtype for
cluster stars ({\it open circles}), template RSGs ({\it filled circles}), and template red giants ({\it filled squares}).
Lines are drawn through the best fits between equivalent width and subtype for the template stars.}
\end{figure}

We can compare these measurements with those of template spectra from \citet{kle86}. As can
be seen in Fig.~\ref{fig:kh}, the bandhead becomes stronger for later spectral types, and it
is generally stronger for supergiants than for giants of a given subtype. These relationships
are quantified in Fig.~\ref{fig:kh_relation}. Considering our measurements, and the trends
evident for the template stars, we conclude that the 14 brightest members of this cluster are
all RSGs with spectral types of M0~I to M4~I. This
is further supported by the fact that the latest spectral types appear to be
redder, as expected (see Fig.~\ref{fig:cmd}). Overall, we estimate an error in the spectral
types of a few subtypes.
\clearpage

\clearpage
The spectrum for star \#15 has weak (but non-zero) CO absorption, and this star is also fainter 
than the 14 brightest stars in the cluster. Its spectrum is more
indicative of a mid-G type supergiant. We suggest that it is a G6~I type based on the 
weak CO features. Given the 
location of the star on the theoretical isochrones (see Fig.~\ref{fig:cmd}) \citep{sch93},
we suggest that it is a cluster member in transition to, or from, the RSG phase.
While such stars are rare, we expect there to be on order one such
star in a cluster with the mass and age we estimate (see below). 

\begin{figure*}
\epsscale{1.6}
\plottwo{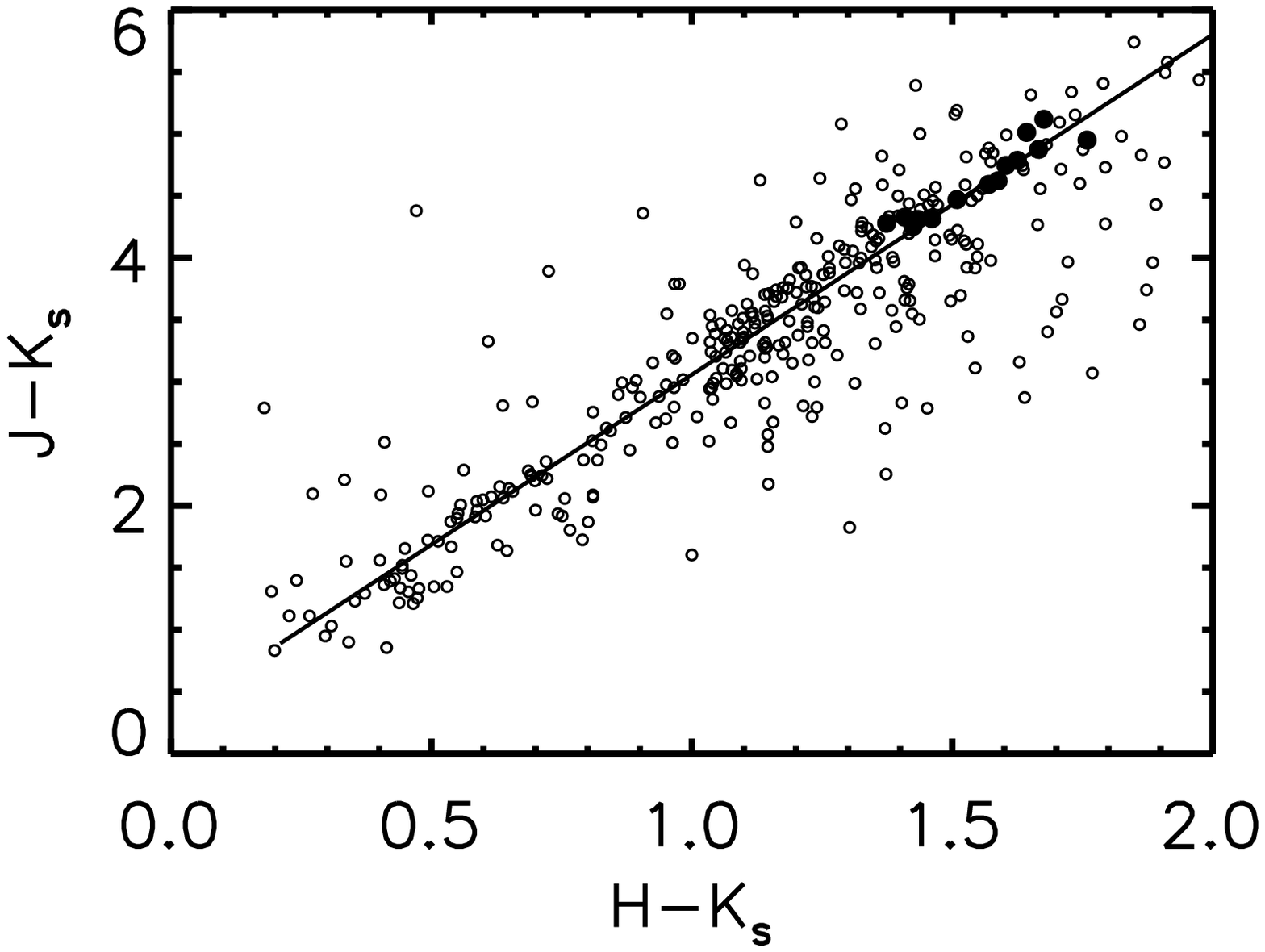}{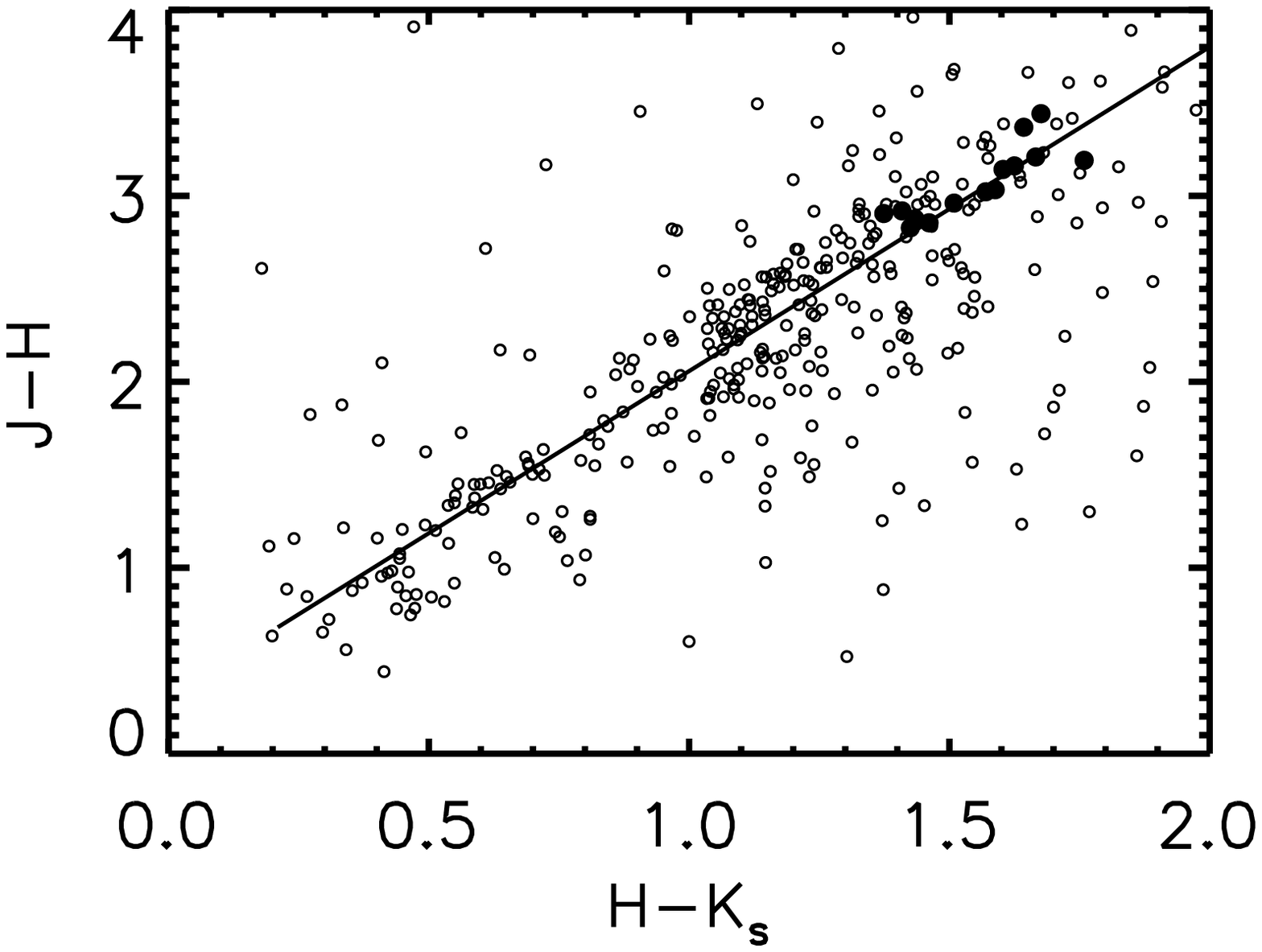}
\caption
{\label{fig:ccJK} {\it 2MASS} color-color diagrams of stars within 3\arcmin\ of the center
of the RSG cluster showing the RSGs ({\it filled circles}) and the fainter field stars 
({\it open circles}). A redenning vector is plotted from the expected 
intrinsic colors of RSGs, assuming the extinction law of \citet{rie89}. The sample has been culled
of stars with reported errors greater than 0.1 magnitudes in {\it K}$_s$. The bulk of the field
stars are not as red as the RSGs, and are thus likely in the foreground.}
\end{figure*}

In any case, the stars fainter than {\it K}$_s$=6.0 in the region are not M supergiants. 
In Fig.~\ref{fig:cmd}, one can identify a cluster of fainter stars in the color-magnitude diagram (CMD) starting at
{\it K}$_s$$\sim$10.
These stars likely represent the main sequence. The expected brightness 
difference between the main sequence stars and the RSGs
is roughly equal to the observed difference. 
The faint cluster stars can also be seen in Fig.~\ref{fig:ccJK}, having colors between
$H-K_s$=1.1 to 1.3; the RSGs are redder in both
colors. The expected color difference between the main sequence stars
and the RSG stars is $H-K_s$$\sim$0.3, in accordance with
the measured difference. 
All the supergiant cluster stars lie on the redenning vector. 
\clearpage

\begin{figure*}
\epsscale{0.6}
\plotone{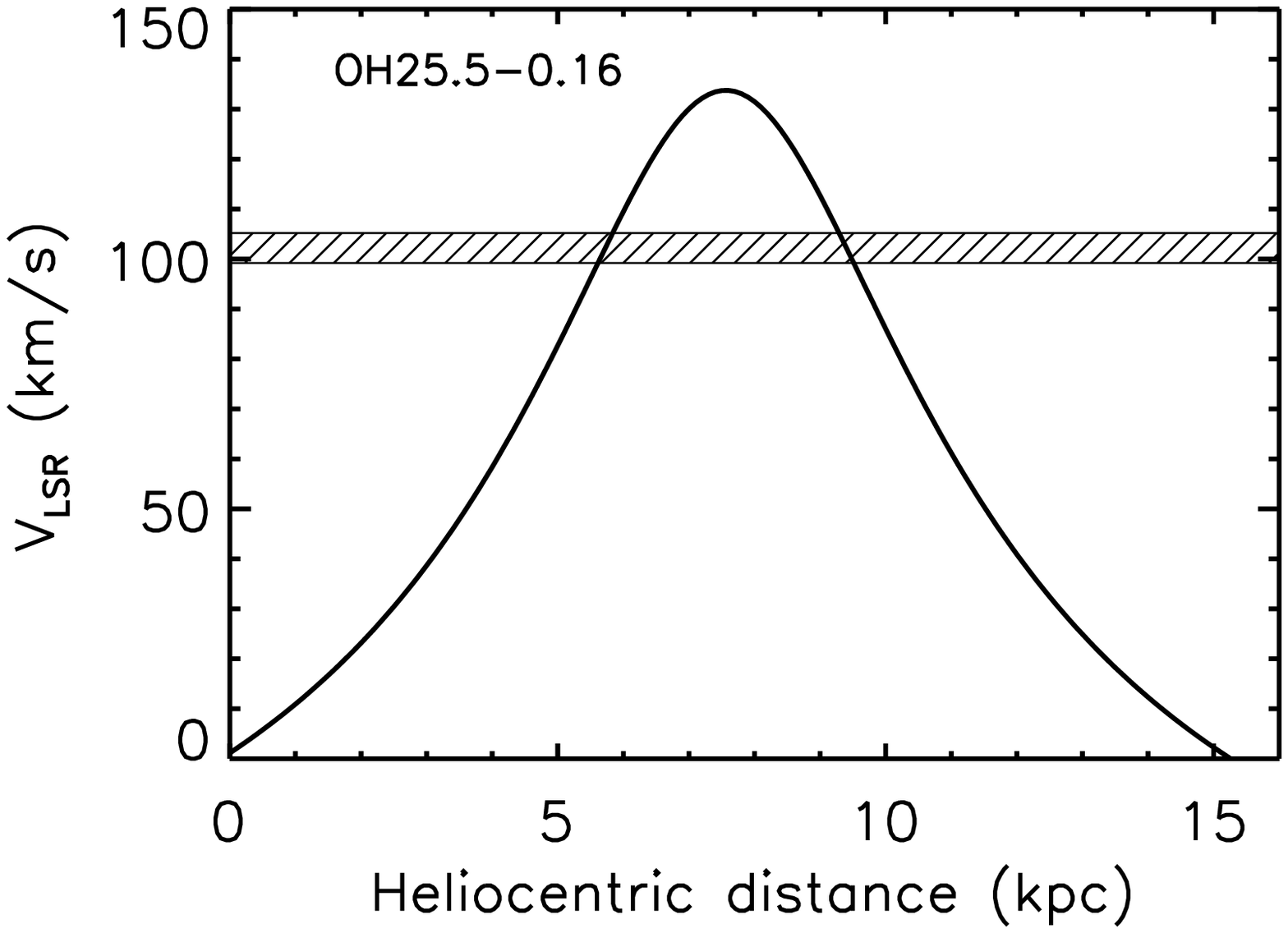}
\caption
{\label{fig:rot} $V_{\rm LSR}$ versus distance along the line of sight to the cluster, assuming the Galactic rotation curve in \citet{bra93}.
The cross-hatched region represents the velocity of the OH~25.5$-$0.16 maser associated with the envelope
of one of the RSGs in the cluster (star \#9). The distance to the cluster is taken to be 5.8~\kpc, corresponding
to the ``near-side'' solution. Note that the ``far-side'' solution would require anomalously high
luminosities for the RSGs.}
\end{figure*}

\subsection{Extinction}
We estimate extinction by adopting average intrinsic colors for M supergiants of H$-$K=0.21 and J$-$K=0.89 \citep{eli85},
and comparing these to the average colors of the RSGs in the cluster. We
identify the difference between the two as the color excess and convert it into \AK$_s$ using 
the relation in \citet{rie89}, finding from E$_{\rm H-K}$: \
\small
\begin{equation}
\AK_s=E_{\rm H-K}/((\lambda_H/\lambda_Ks)^{-1.53}-1)=(1.55-0.21)/((1.662/2.159)^{-1.53}-1)=2.73,
\end{equation}
and from E$_{\rm J-K}$:
\begin{equation}
\AK_s=E_{\rm J-K}/((\lambda_J/\lambda_Ks)^{-1.53}-1)=(4.62-0.89)/((1.235/2.159)^{-1.53}-1)=2.76. 
\end{equation}
\normalsize
We adopt the average of these two inferred extinctions, \AK=2.74, throughout the rest of this paper.

\subsection{Distance}
OH~25.25$-$0.16 is an OH maser source centered about 5\arcsec\ north of the
cluster center, at RA(2000)=18$^h$37$^m$58.27$^s$, Dec(2000)=$-$06$\arcdeg$52$\arcmin$28$\farcs$8. 
\citet{blo94} quote a ``typical'' positional uncertainty for their sample of 4$\arcsec$; it is unclear
if this uncertainty applies to the position of OH~25.25$-$0.16. 
The maser is roughly equidistant, d$\sim$9$\arcsec$, from stars~\#9, \#13, and \#15 (see Fig.~\ref{fig:cont}). 
These types of masers are often produced in 
the stellar winds of latest (M4-M5) RSGs \citep{lew91}, i.e. S~Per, VX~Sgr, NML~Cyg
and VY~CMa; see \citet{hum75} for a review. We believe that the maser is likely produced
by either star~\#9 or \#13, as \#15 is too warm to be associated with such a maser. OH masers
and RSGs are relatively rare, and we estimate a random probability of 10$^{-10}$ 
of a chance alignment having the observed separation. 
Unlike most maser sources, this has only one
velocity peak, at 102.2~\kms. If we take this as the systemic velocity for
the maser, and associate the soure with the cluster, then we can estimate a
distance to the cluster, assuming the Galactic rotation curve in \citet{bra93}. Doing so
yields a distance of 5.8~\kpc\ (Fig.~\ref{fig:rot}). This ``near-side'' distance places the RSG
stars exactly at the luminosities expected from the isochrones (Fig.~\ref{fig:cmd}). If the 
stars were at the far side solution implied by a fit to the Galactic rotation
curve, they would then have L$\sim10^6$~\Lsun, a factor of four higher than for typical RSGs.

\subsection{Luminosities}
\begin{figure}
\epsscale{1.2}
\plotone{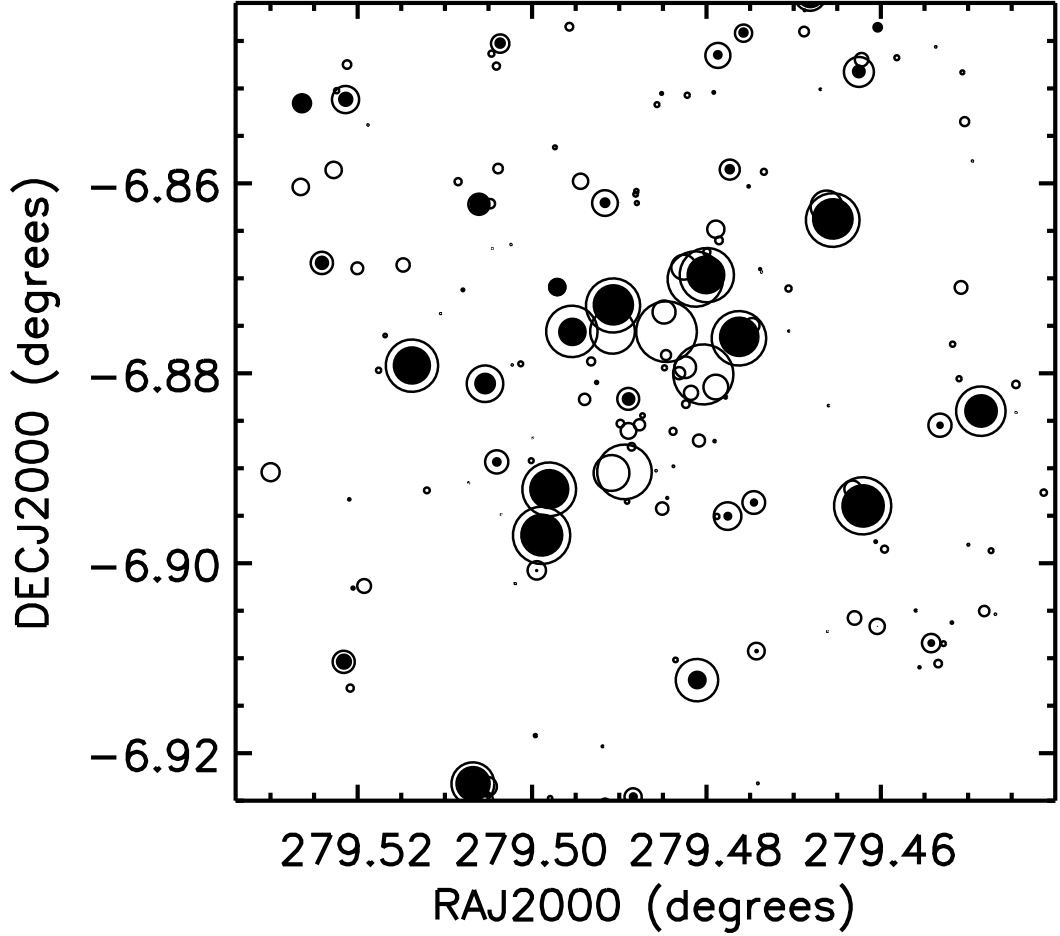}
\caption
{\label{fig:spitzer} Plot of supergiant cluster stars from {\it 2MASS} ({\it open circles}) and 
{\it Spitzer}/GLIMPSE at 3.6~\micron\ ({\it filled circles}). Symbol
sizes scale with measured flux.}
\end{figure}

\begin{figure}
\epsscale{1.1}
\plotone{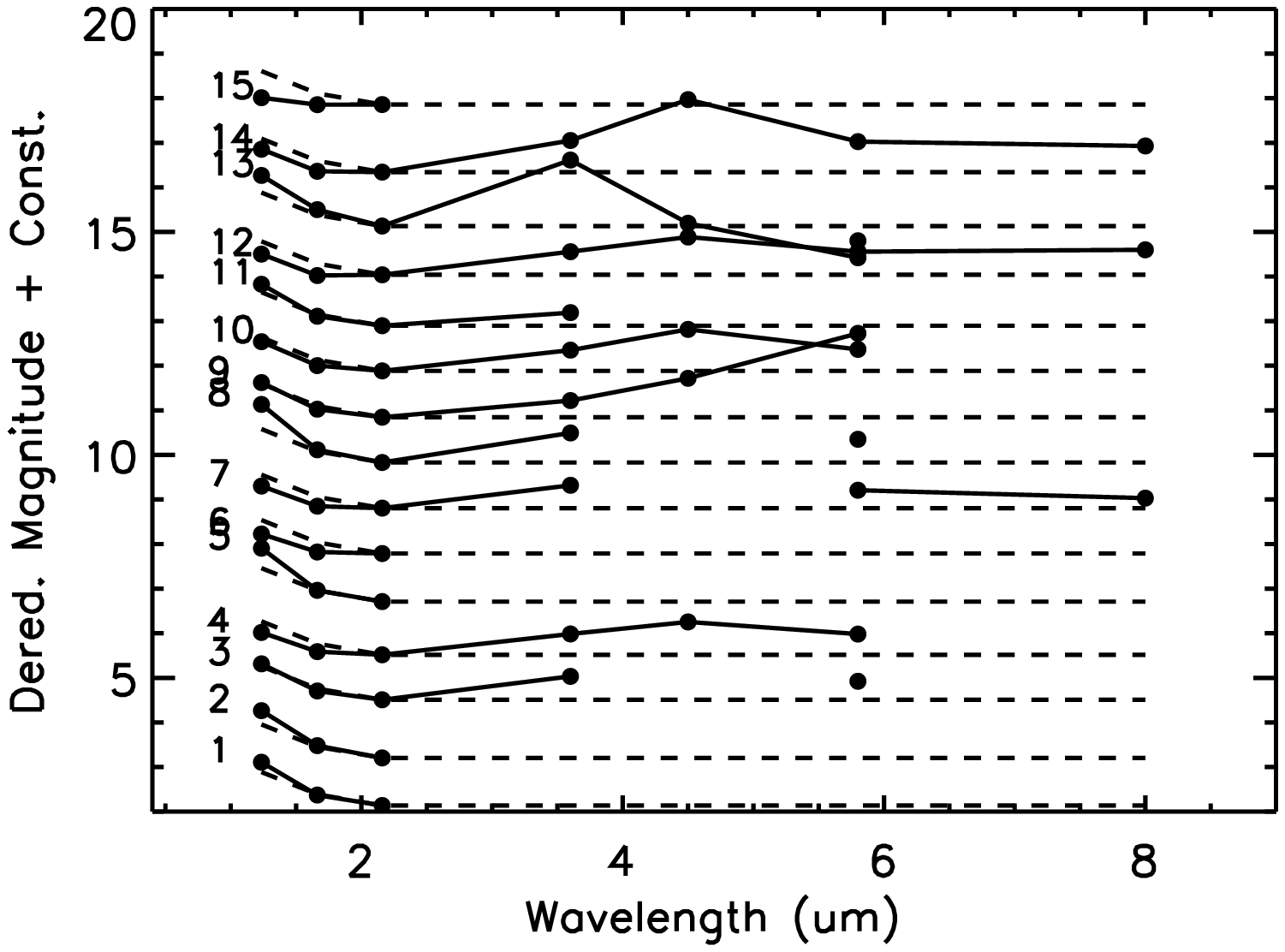}
\caption
{\label{fig:mags} Plot of dereddened magnitudes for cluster stars ({\it filled circles, solid lines}). The values have been
shifted by their ID number for presentation purposes. The intrinsic magnitudes of
RSG stars are overplotted ({\it dashed lines}).
The upward trend of the
magnitudes toward shorter wavelengths is expected for such cool stars on
the magnitude system. Note that the differences between the observations and model are
closely related to the inferred spectral types. For instance, the RSG model
significantly underpredicts the flux at short wavelengths for star \#15, a G6~I star, as
would be expected.}
\end{figure}

We can estimate the total luminosity of each star by integrating its spectral energy
distribution. Being relatively cool, RSGs emit most of their radiation near
1~\micron, although mid-infrared flux may contribute significantly in the case of photon
reprocessing by circumstellar dust. 
Most of the RSG stars can be identified in observations from the {\it Spitzer/GLIMPSE} survey 
(see Fig.~\ref{fig:spitzer}), although several cannot because such a crowded field poses problems
for the photometry extraction process. 
Fig.~\ref{fig:mags} shows plots of the spectral energy distributions for the RSGs. 
As expected for cool stars, the magnitudes trend higher (fainter) at the shortest wavelengths. 

\begin{figure*}
\plotone{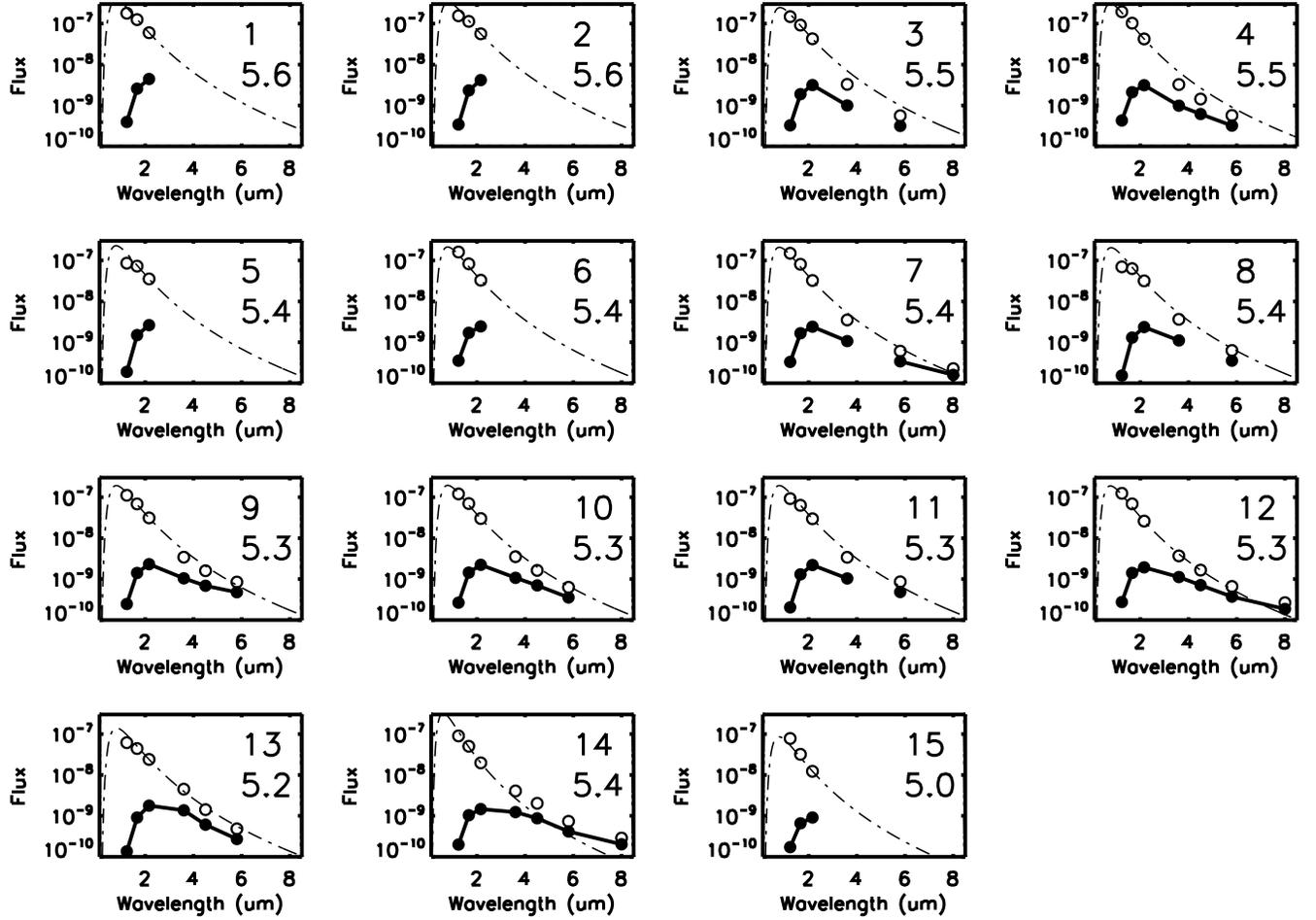}
\caption
{\label{fig:blackbody} Plots of photometry from {\it 2MASS} and {\it Spitzer}/GLIMPSE surveys for the
14 supergiants in this paper. The {\it solid lines} are drawn through the observed fluxes ({\it filled circles}),
in units of $ergs~s^{-1}~cm^{-2}~\micron^{-1}$.
The {\it dashed lines} are drawn for blackbodies fitting the dereddened photometry ({\it open circles}) and having temperatures
appropriate for the spectral types of the stars. The logs of the integrated 
luminosities, in solar units, under the blackbody curve are given in the
legend.}
\end{figure*}

Fig.~\ref{fig:blackbody} shows blackbody fits to the dereddened fluxes from the {\it 2MASS} and {\it Spitzer} data. 
Each blackbody has been assigned the appropriate temperature corresponding to the spectral type \citep{lev05}, although
there may be evidence of a long wavelength excess in some cases, i.e. \#9, \#11 and \#12. We computed
the integrated luminosity under the blackbody curves, and the log of these values, in solar
units, are plotted in the legends of the figure panels. All the luminosities are consistent with
what would be expected for supergiants. Taken at face value, they suggest M$_{\rm bol}\sim -8~to~-9$,
with correspondingly high implied initial masses of \Minit=15 to 25~\Msun.

\clearpage

\section{Discussion}

The cluster is extraordinary for its membership of 14 RSGs, more than
any other cluster in the Galaxy. We show below that this implies a cluster  mass 
of at least 20,000~\Msun, comparable to the masses of the most massive young clusters
in the Galaxy. The implied age leads us to believe that the cluster is ripe
for supernovae activity. Indeed, we claim that the present-day supernovae rate is consistent
with the presence of high energy objects in the field of the cluster. 

Relatively little is known about this cluster. It is designated as candidate cluster \#122 in
the list of \citet{dut03a}, who note that it is in/near the G25.253$-$0.150 and 
W42 regions, that it appears to be an infrared cluster, and that it is a loose collection
of stars spanning approximately 4\arcmin\ on the sky. The cluster is 
about 7\arcmin\ to the southwest of the massive stars identified by \citet{blu00} in
W42 (see Fig.~\ref{fig:msx}). Our distance estimate places the RSG cluster within
the rather large range of estimates for the distance of W42: 
2.2~\kpc, 3.7~\kpc, 6.0~\kpc, and 9.3~\kpc\ \citep{les85,chu90,blu00,cro03}.
The bulk of the information known about this cluster is determined in this paper and described in the following sections. 

The remainder of this section of the paper is devoted to establishing the
characteristics of the RSG cluster in the context of massive young
stellar clusters in the Galaxy. We find that the cluster is extraordinary for its
content of RSGs, and readily interpreted as a very massive cluster at
an age ($\sim$10~\Myr) when it produces an extraordinary number of RSGs.

\subsection{Cluster Mass and Age}
The cluster mass can be estimated by fitting a reasonable initial mass function (IMF)
through the number of stars in the cluster with known initial masses, in this case, the RSGs. 
We can relate the RSG luminosities to initial masses by using stellar
evolution models, age estimates, and assumptions about the metallicities and stellar mass-loss rates. 
To this end, we use a Monte Carlo simulation code that randomly draws
initial stellar masses from a uniform distribution constrained by a Salpeter IMF \citep{sal55}, and
truncated at 0.8~\Msun\ and 150~\Msun\ \citep{fig05a}. We then choose an
isochrone to convert initial mass
to present day mass, temperature, luminosity, and absolute magnitude. For the figures in
this paper, we choose the Geneva models with solar metallicity and canonical mass-loss rates. 
We convert to apparent magnitudes by adding the observed extinction and the distance modulus. 
Finally, we identify all the supergiant stars ({\it L}/\Lsun$>10^{4.5}$) as falling into one of
three categories, RSG (T$<$4500~K), YSG (4500~K$<$T$<$10000~K), or BSG (T$>$10000~K). 

\begin{figure*}
\epsscale{1.4}
\plottwo{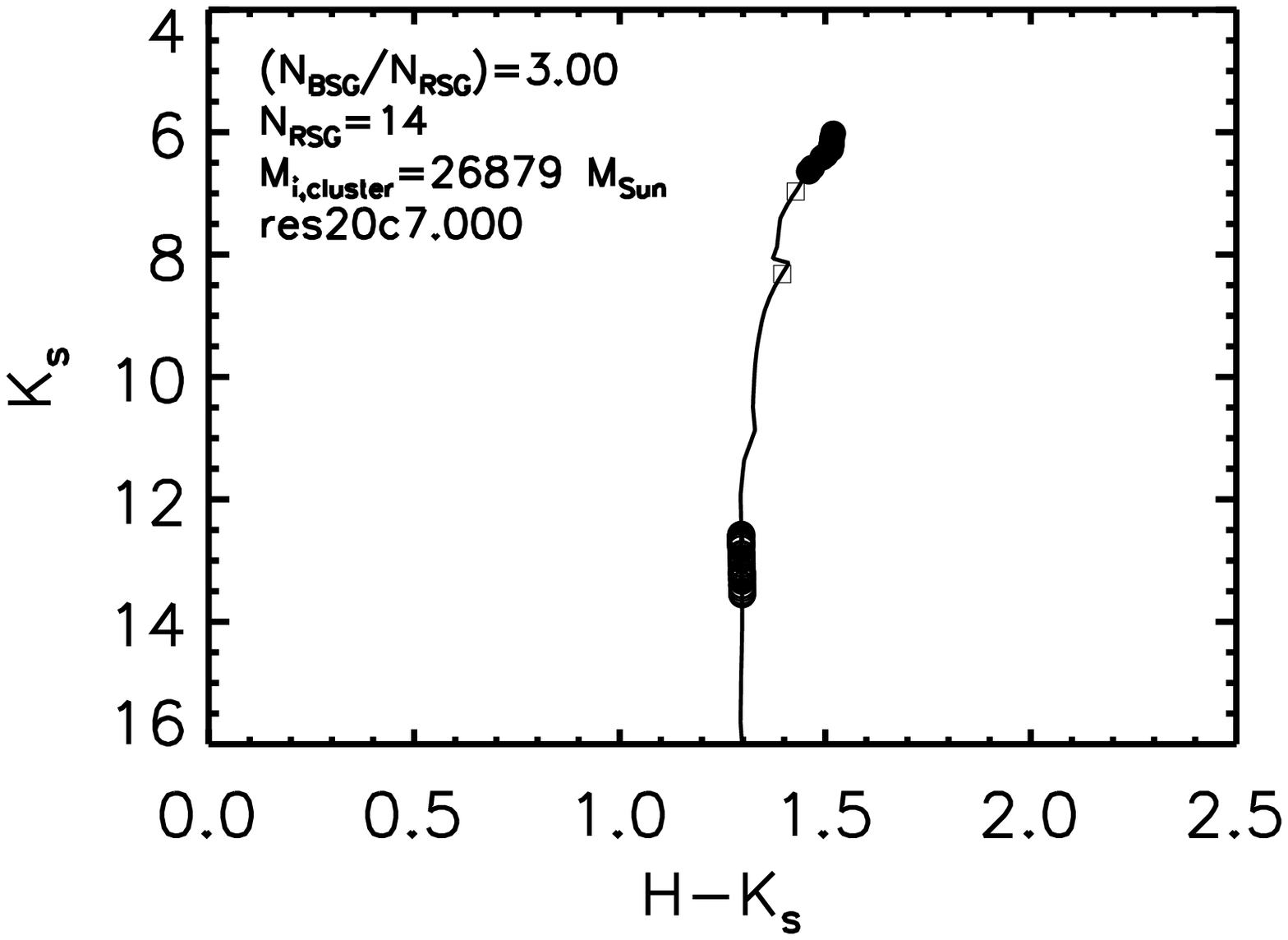}{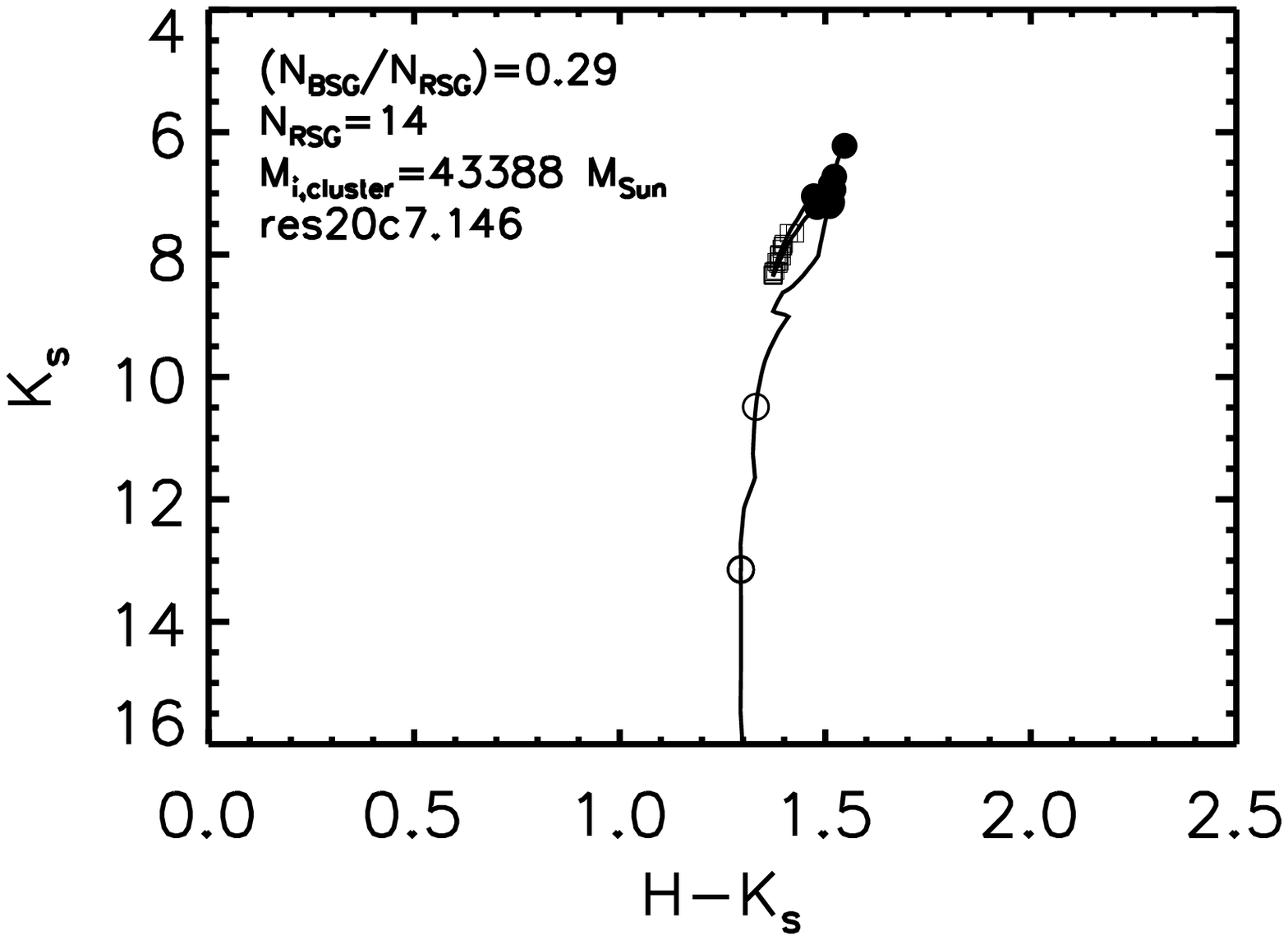}
\caption
{\label{fig:cmd_monte} Sample CMDs from a Monte Carlo simulation for ages of 10~\Myr\ ({\it left})
and 14~\Myr\ ({\it right}), assuming the non-rotating Geneva models with solar metallicity and
the canonical mass-loss rates. The symbols at the upper portion of the diagram are RSGs ({\it filled circles})
and those near the bottom are BSGs ({\it open circles}). YSGs
are designated by {\it open squares}.}
\end{figure*}

\begin{figure*}
\plottwo{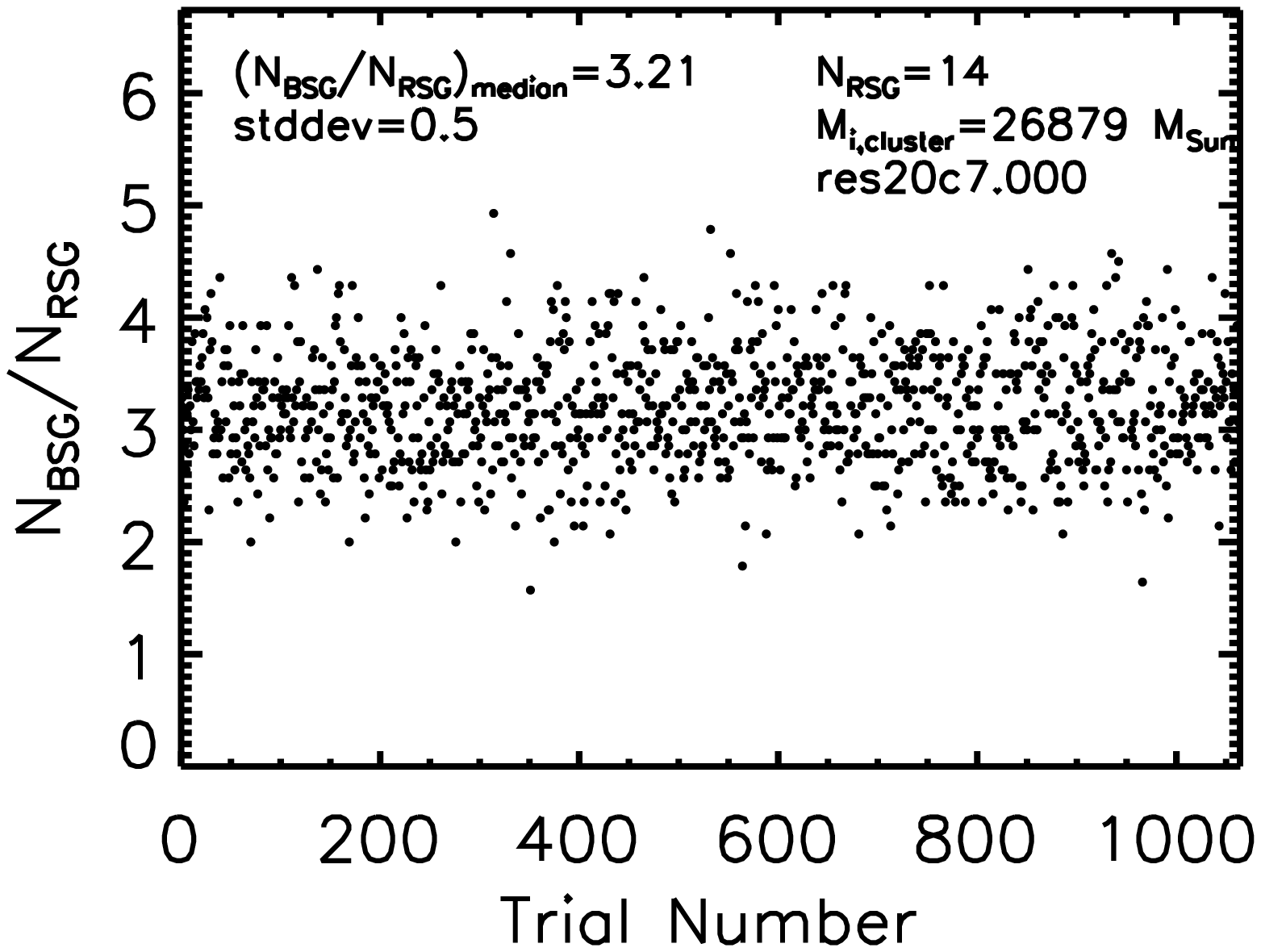}{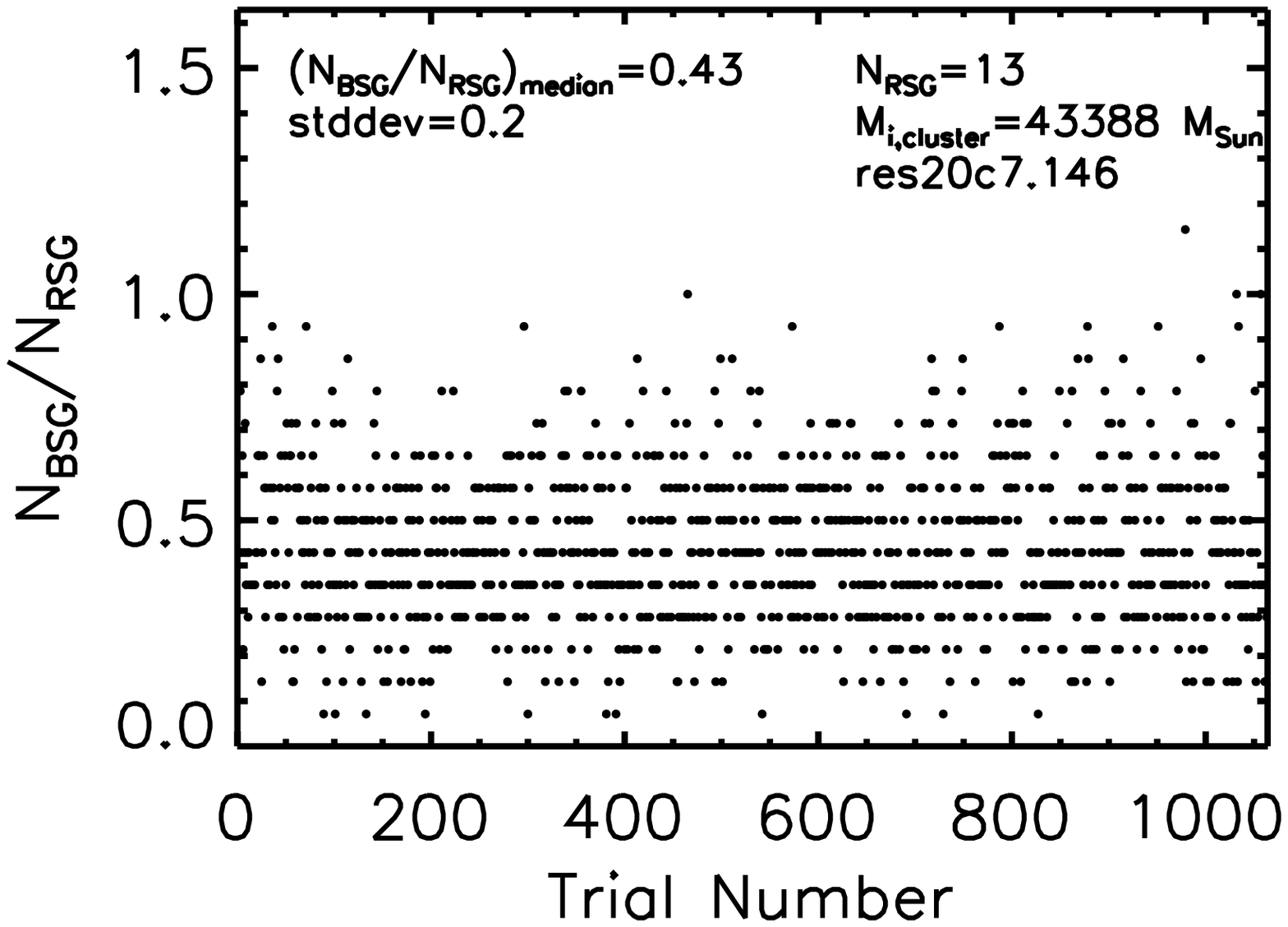}
\caption
{\label{fig:ratio_monte} The ratio of BSGs to RSGs in sample trials from the Monte
Carlo simulation. The cluster masses are randomly drawn from a distribution constrained to
follow a Salpeter IMF such that the median number of RSGs is equal to 14, the observed
value in the cluster. The plots are for ages of 10~\Myr\ ({\it left})
and 14~\Myr\ ({\it right}). The legend gives the median value of the BSG to RSG ratio, the
median number of RSGs, the total estimated initial cluster mass, and the model isochrone
file name.}
\end{figure*}

\begin{figure*}
\plottwo{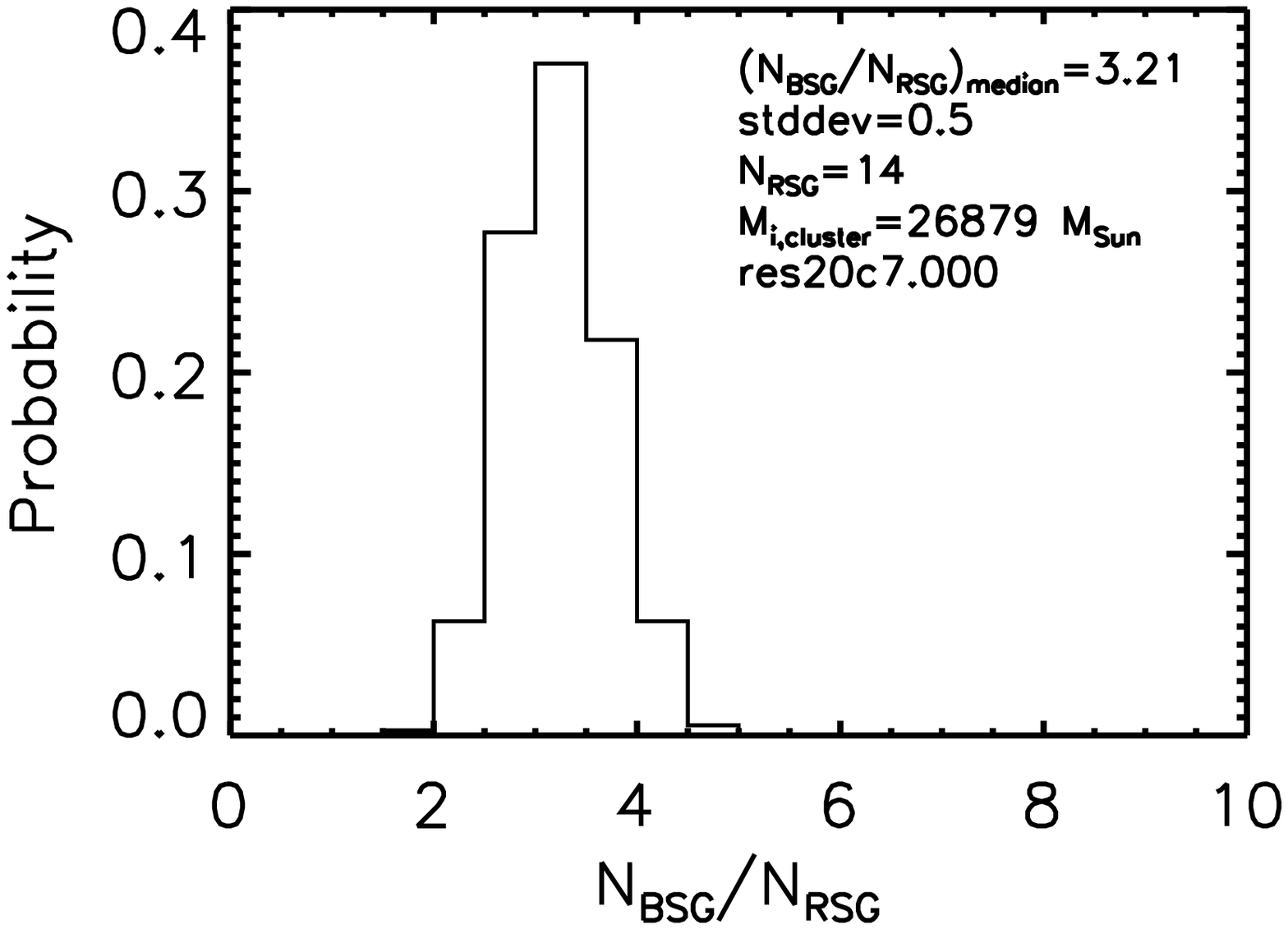}{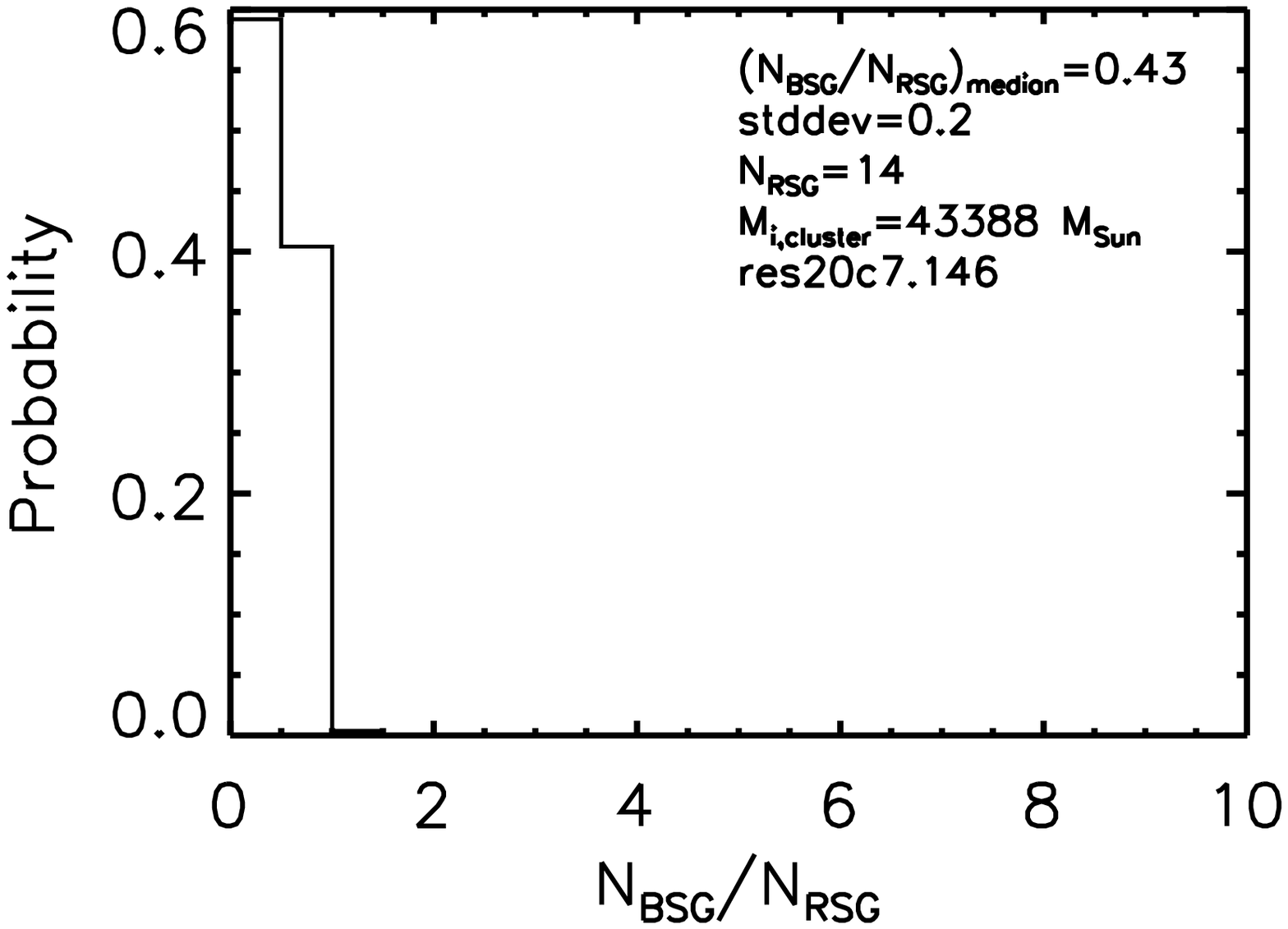}
\caption
{\label{fig:hist_monte} Histogram of the sample defined in Fig.~\ref{fig:ratio_monte}. 
The plots are for ages of 10~\Myr\ ({\it left})
and 14~\Myr\ ({\it right}).}
\end{figure*}

\begin{figure*}
\plottwo{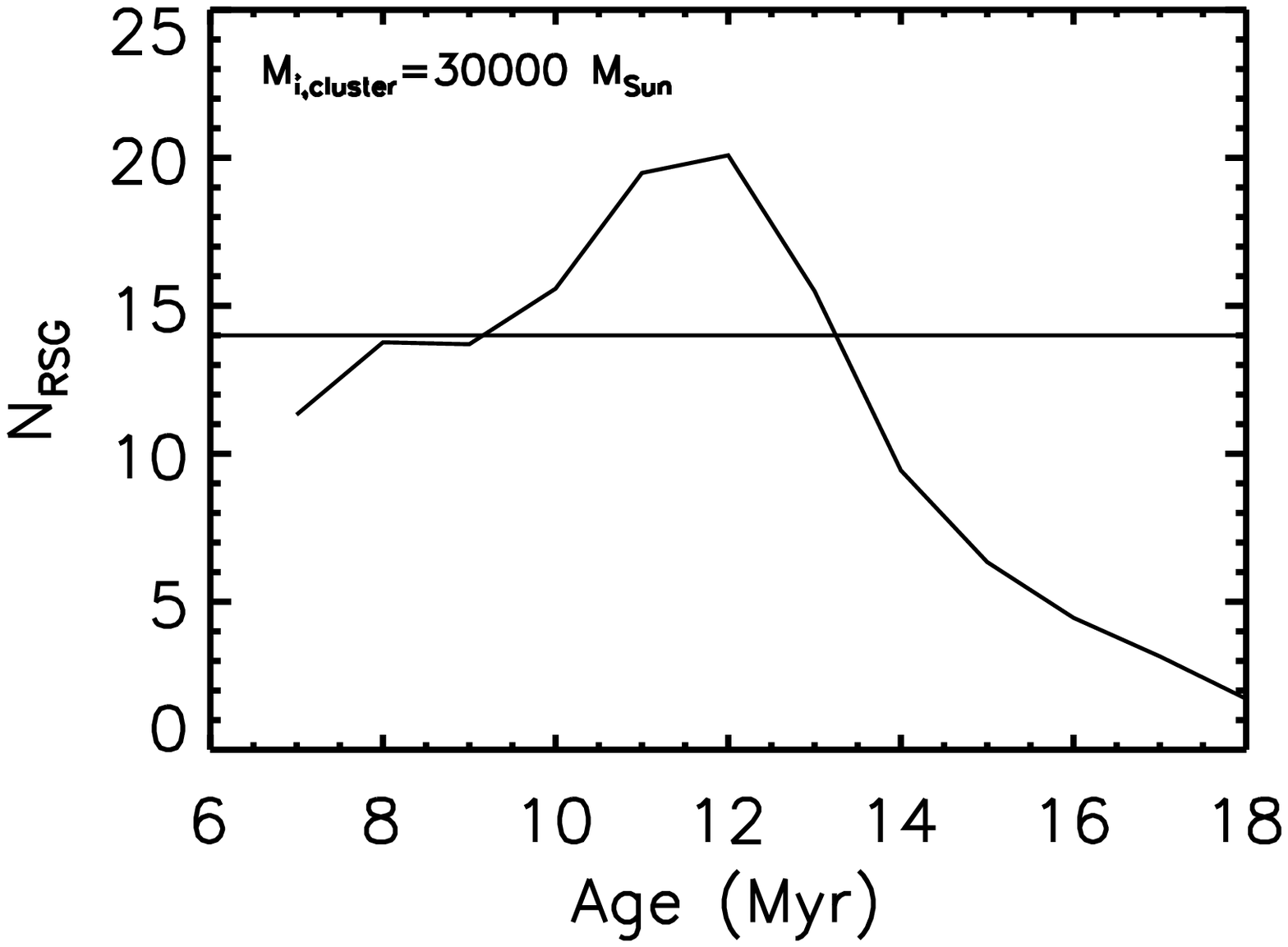}{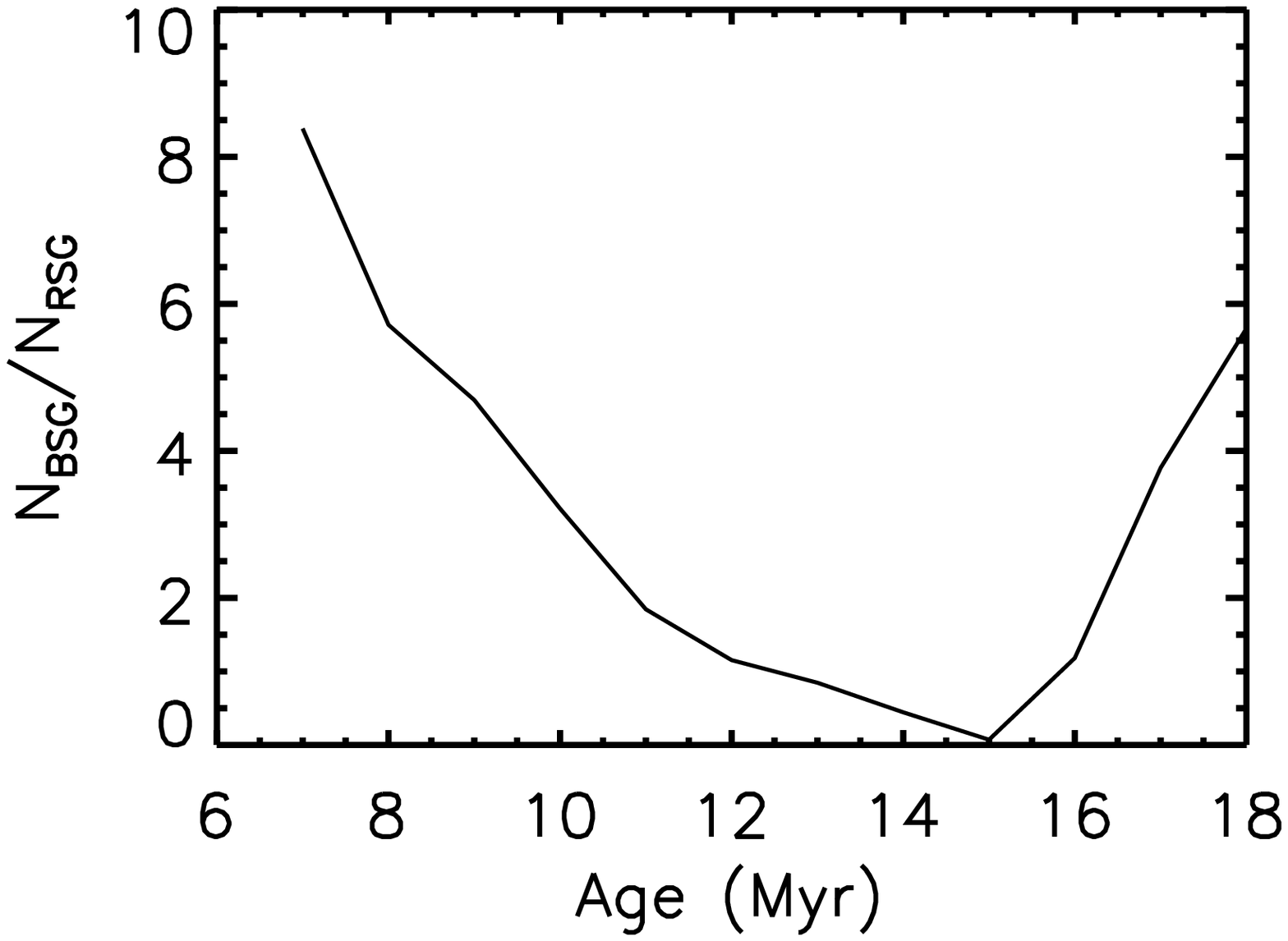}
\caption
{\label{fig:age_monte} The progression of the number of RSGs ({\it left}) and the ratio
of BSG to RSGs ({\it right}) as functions of age for an initial cluster mass of 30,000~\Msun,
assuming the non-rotating Geneva models with solar metallicity and
the canonical mass-loss rates. The plot of RSGs can be scaled vertically as a 
linear function of initial cluster mass to match the number of observed RSGs indicated
by the horizontal line. }
\end{figure*}

\begin{figure}
\epsscale{1.1}
\plotone{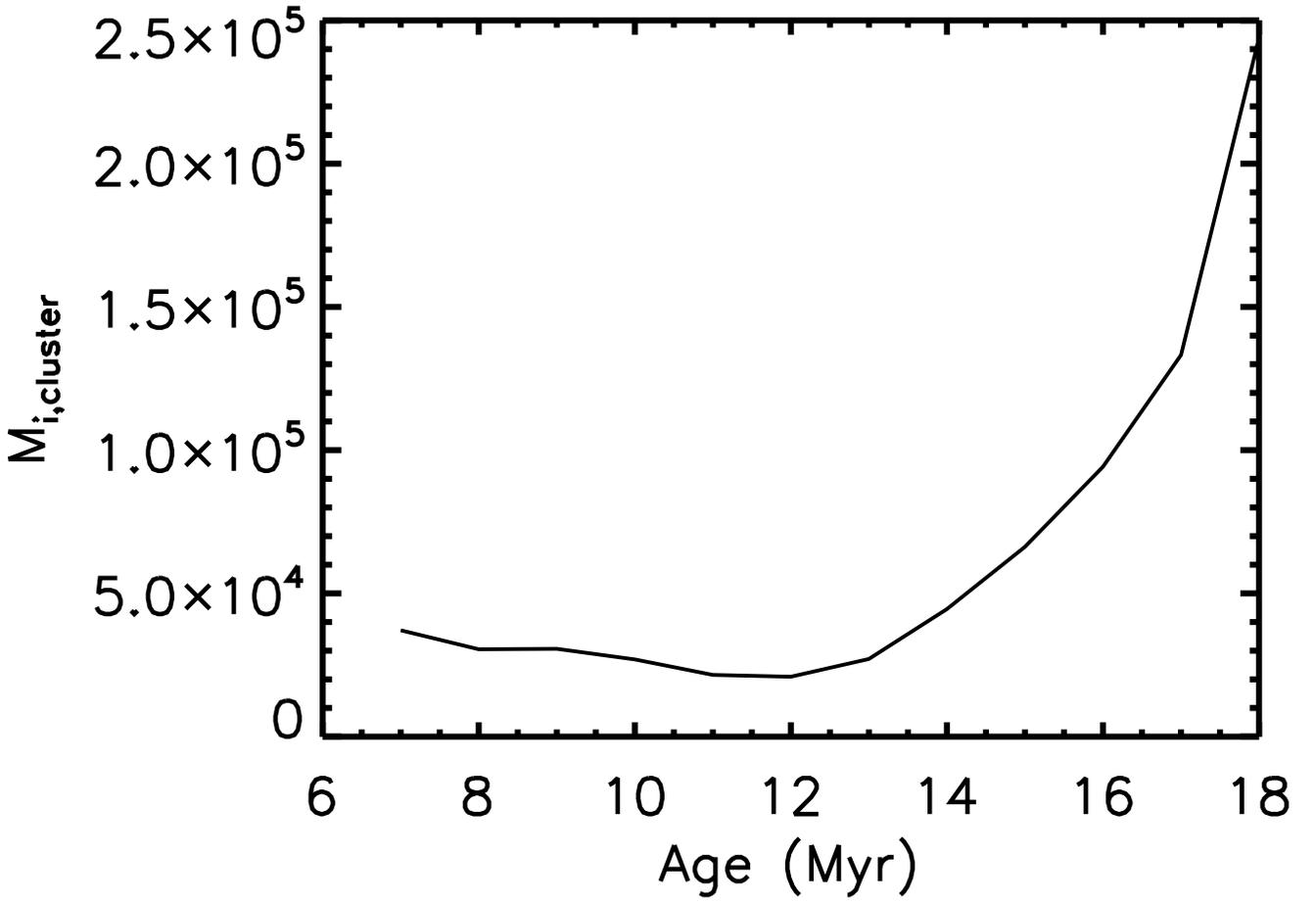}
\caption
{\label{fig:mass_monte} Implied initial cluster mass as a function of time, assuming a
cluster having 14 RSGs and the Geneva models with solar metallicity.}
\end{figure}

One set of sample clusters is shown in Fig.~\ref{fig:cmd_monte} for ages of 10~\Myr\ and 14~\Myr. 
As expected, we see that the number of BSGs to RSGs is
lower for the older cluster. We are motivated to examine the behavior of the BSG to RSG ratio because it 
provides a critical test for stellar evolution
models, and it determines the spectral appearance of integrated
light from starburst populations \citep{lan95,kip90,mae00,mae01}. 
The ratio for each trial is shown in Fig.~\ref{fig:ratio_monte} for ages of 10~\Myr\ and 14~\Myr. These
plots demonstrate the statistical scatter induced by randomly sampling 
the IMF; a histogram of these values is shown in Fig.~\ref{fig:hist_monte}.
We can analyze the Monte Carlo results by examining statistical
properties of the whole sample. For instance, Fig.~\ref{fig:age_monte}
shows the number of RSGs as a function of age for a cluster with initial mass of 30,000~\Msun,
and the associated ratio of BSGs to RSGs, both computed as the median over the set
of sample clusters. We see that the number of RSGs peaks at $\approx$12~\Myr, and
the ratio of BSGs to RSGs is steadily decreasing as a function of age. The implied initial
cluster mass is shown in Fig.~\ref{fig:mass_monte}, assuming a cluster having 14 RSGs.

\begin{figure}
\epsscale{1.15}
\plotone{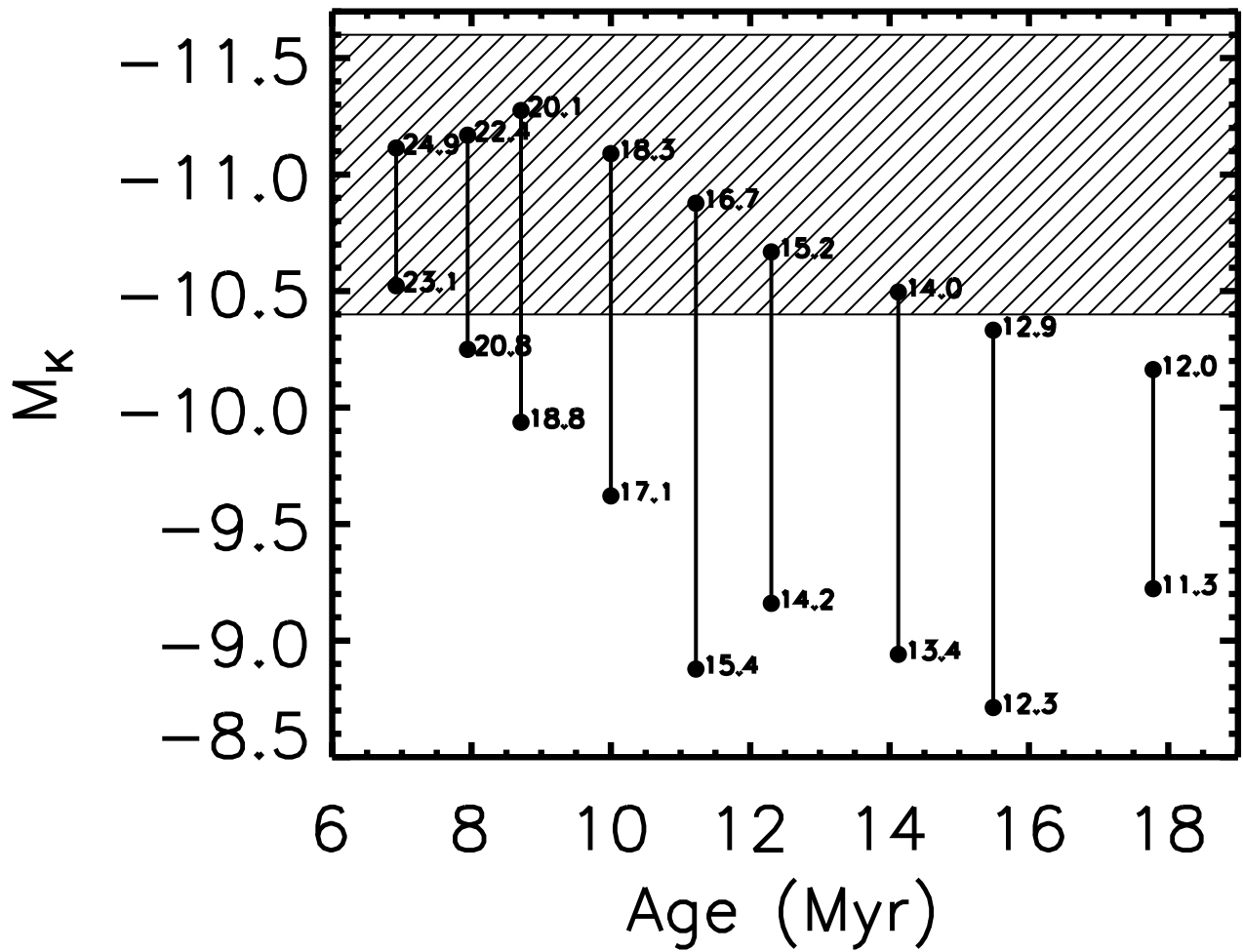}
\caption
{\label{fig:model} $M_{\rm K}$ versus cluster age for the most luminous and least luminous RSGs in
a cluster, according to the Geneva models with solar metallicity and the canonical mass-loss rates. 
Initial stellar masses are plotted as data labels. Stars in the RSG cluster have 
$M_{\rm K}\sim-10.5~to~-11.5$, as shown in the cross-hatched region. Variations in metallicity or
mass-loss rate do not significantly alter the trend that the observations are best fit
by a relatively young cluster.}
\end{figure}

The cluster mass and stellar luminosities depend on age. Fig.~\ref{fig:model} shows a
plot of M$_{\rm K}$ versus cluster age for the most luminous and least luminous RSGs in
a cluster demonstrating this behavior. From inspection of this figure, it appears that the
high luminosities for the stars in the RSG cluster are consistent with a relatively young
age ($\sim$7-12\Myr). 

\clearpage

\clearpage
Several things are immediately apparent from these models. First, they 
require that the initial cluster be quite massive, at least 20,000~\Msun, and perhaps
as high as 100,000~\Msun, where the actual value primarily depends on age. 
Second, there is a large brightness gap
between the RSGs and the next brightest group of stars in the CMD, matching the
observations. Lastly, the BSG/RSG ratio can vary just from statistical sampling effects. 
For instance, it would be quite plausible to observe a cluster with $\approx\pm$20\% of the median ratio,
even if the stellar evolution models are perfectly accurate for that cluster. Of course, this
effect scales strongly with the number of stars in the cluster, i.e. the standard deviation
of the histogram scales roughly inverse linearly with the total initial cluster mass. 

We also examine trends with metallicity. We find that twice-solar metallicity model clusters
produce more RSGs, i.e.\ the initial cluster masses might be only $\sim$20,000~\Msun\
for all ages, whereas the BSG to RSG ratio is roughly half of that for solar metallicity.

Our large cluster mass estimate depends critically on the large number of RSGs in the cluster. There are two
possibilities that could reduce the estimated total mass of the cluster: binarity/multiplicity and non-coevality.
Some objects could be  multiple (although even then, the large gap in Ks between the RSGs and the main sequence 
indicates that most of the bright sources would be RSGs, but not as extreme). Without higher resolution
spectroscopy, it is difficult to asses the impact of this possibility. It is also
possible that the bright stars were formed in several bursts, or via continous star 
formation, over a substantial timescale. Fig.~\ref{fig:model} implies that 
the RSG progenitors could come from initial stellar masses spanning a range of 25 to 18.5~\Msun\ for 
an age range spanning 2~\Myr. While this age spread is typical for errors in estimating a cluster
age, it would imply an extended star formation episode. 

\clearpage

\clearpage

\subsection{High Energy Objects Near the RSG Cluster}
The field surrounding the cluster contains a number of high energy sources, and 
Figs.~\ref{fig:cont} and \ref{fig:msx} show their spatial relationships. In this section,
we interpret these sources as potential indicators of recent supernova activity in the cluster.

The {\it Advanced Satellite for Cosmology and Astrophysics} ({\it ASCA}) mission 
identified several sources in the field \citep{bam03}. The center of the error circle for 
AX~J1838.0$-$0655 is located to the 
south of the cluster center, and its flux in the 0.7$-$10 keV energy band is 1.1(10$^{-11})~ergs~cm^{-2}~s^{-1}$. 
The visual extinction to the source is estimated to be A$_{V}$=36.3 \citep{mal05}, 
somewhat higher than the value we find for the RSG cluster (A$_{V}$=2.74/0.112=24.5).
Another x-ray source, AX~J1837.5$-$0653, is located to the west by 5\arcmin\ \citep{bam03}. 

The {\it INTErnational Gamma-Ray Astrophysics Laboratory} ({\it INTEGRAL}) gamma-ray observatory 
also observed a source in the field \citep{mal05}, to the south and east of the cluster center. 
Its flux in the 20-300~keV energy range is 9(10$^{-11}$)~{\it erg~cm$^{-2}$~s$^{-1}$}. Given the
relatively large error circle for this source, it is possible that the source of the flux is located
somewhere in the cluster. 

The {\it High Energy Spectroscopic System} ({\it HESS}) telescopes detected high energy gamma
rays (E$>$100~{\it GeV}) from HESS~J1837$–$069, which is located to the southwest of the cluster (see Fig.~\ref{fig:msx}), at 
R.A.(2000)= 18$^h$37$^m$42.7$^s$ and Dec(2000)=$-$6$\arcdeg$55$\arcmin$39$\farcs$0, 
with a size of 7\arcmin\ and a positional uncertainty 
of 1-2\arcmin\ \citep{aha05}. We estimate a random chance alignment of the cluster and
source of $\approx$0.05\%. 
The {\it HESS} source has a flux of 9(10$^{-12}$)~{\it erg~cm$^{-2}$~s$^{-1}$} above 200~{\it GeV}, or 13.4\% of
that from the Crab nebula. Assuming that the {\it HESS} source is associated with the
RSG cluster, then it has a luminosity of (5800/1930)$\times$13.4\%=120\% of that of
the Crab. This would suggest that a recent supernovae in the cluster is responsible
for this source. \citet{aha05} note that
the {\it HESS} object is likely a supernovae remnant or pulsar wind driven nebula. 

We believe that the high energy sources are physically related to the RSG
cluster. Indeed, there is precedence for associating extended high energy sources with a
cluster of massive stars, i.e.\ in CYG~OB2 \citep{aha02}. \citet{mal05} note that 
AX J1838.0$-$0655 is very near to a bright mid-infrared source detected by MSX 
(MSX6C G025.2454$-$00.1885), but that the latter is likely too bright to be consistent
with the large redenning inferred for the former. We argue the opposite. The MSX source
is the cluster of RSGs, and it is as bright as expected for such an object that 
is heavily extincted. Therefore, we associate the high energy objects with the cluster. 
If we associate the {\it ASCA} source (AX~J1838.0$-$0655), the INTEGRAL source, 
and the {\it HESS} source, with
the same object, then the total x-ray, gamma ray, and TeV 
luminosities are 4.4, 36, and 3.6(10$^{34})~ergs~s^{-1}$ \citep{mal05},
assuming d=5,800~{\it pc}. 

There are several other high energy sources near the cluster that are either not
related, or whose association with the cluster is uncertain. 
PSR~J1837$-$06 is located 11\arcmin\ to the west of the cluster, at 
RA=18$^h$37$^m$14.65$^s$, DEC=$-$6$\arcdeg$53$\arcmin$2$\farcs$1 (J2000). 
It has a period of 1.9~$s$ and a
distance of $\sim$5 kpc, a spin down age $\tau$=39~\Myr, where the spin down age is defined to be $P/2\dot {P}$, 
and a spin down flux 
of $1.5(10^{-15})~erg~cm^{-2}~s^{-1}$ \citep{mal05} or 2.5(10$^{26})~ergs~s^{-1}$ at 5800 pc.
The spin down rate is $\dot {P}$=7.7(10$^{-16}$)~$s~s^{-1}$. The magnetic
field strength is 3.2(10$^{19}$)($P\dot {P}$)$^{0.5}$, or 1.2$(10^{12})$~{\it G} 
(not enough to be a magnetar, $\sim$10$^{14}$~{\it G}).  
Note that this PSR is likely NOT associated with the {\it HESS} source \citep{mal05}, and
it is quite a distance from the cluster.
The hard x-ray source, G25.5+0.0 is nearby and thought to be a supernova remnant \citep{bam03}. Given
its projected distance from the cluster, it is unclear whether it is related. 

\subsection{Radio/IR Objects Near the RSG Cluster}
\begin{figure}
\epsscale{1.15}
\plotone{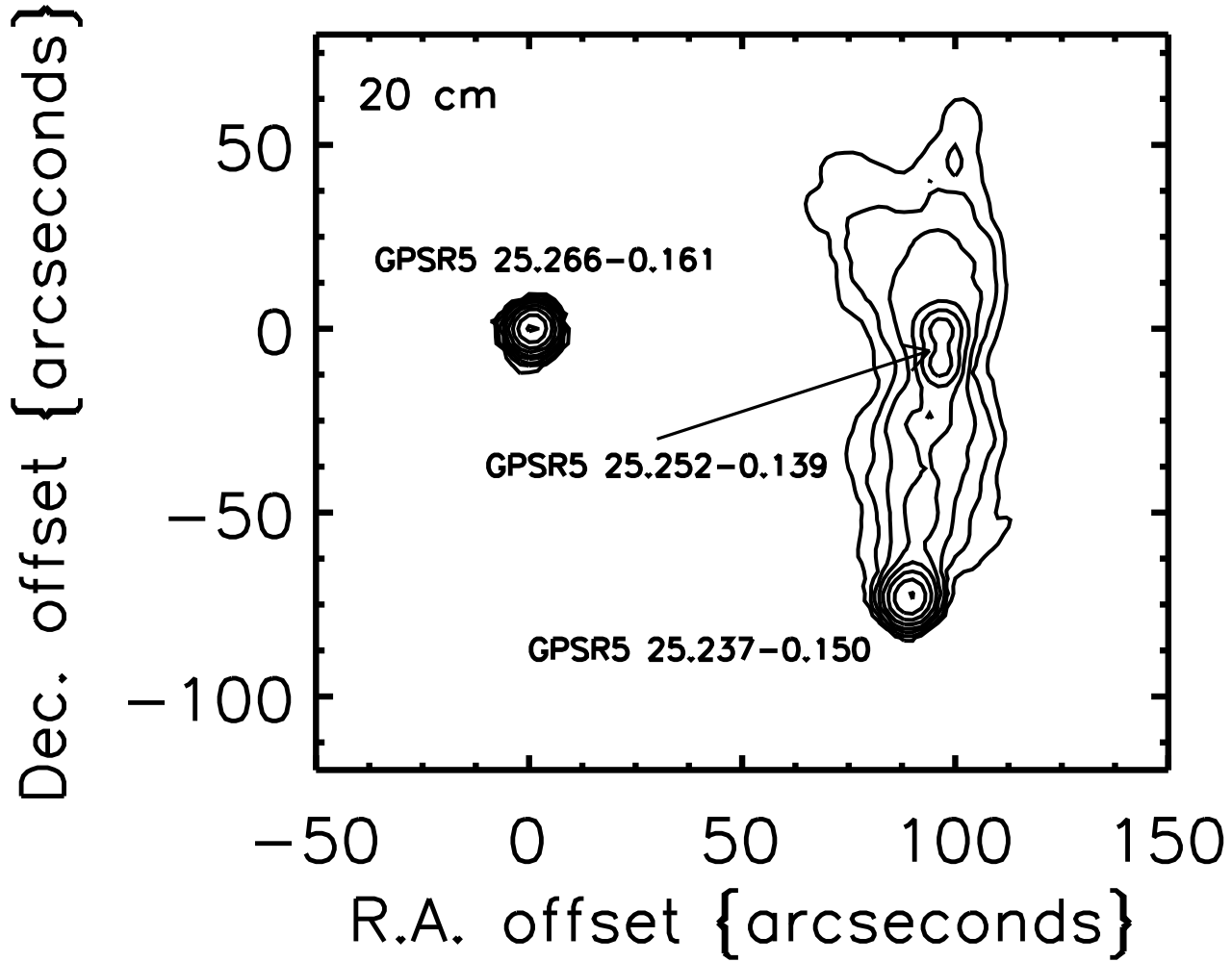}
\caption
{\label{fig:contour_20cm} Contour plot of radio flux at 20~cm \citep{whi05,hel05} within the 
cluster field (see Fig.~\ref{fig:cont}). The contours are drawn at flux levels, in mJy per beam:
2, 4, 8, 16, 32, 64, 128, 256, 512, 1024.
The beam size is $6\farcs 2\times5\farcs 4$, with the long axis along the north-south direction.}
\end{figure}

There are a number of radio objects in the field, as shown in Fig.~\ref{fig:cont}. 
\citet{alt79} identify G025.252$-$0.150 as a bright 6.1~cm source (0.81~Jy) 
centered on the cluster field. 
\citet{hel89a} resolve this source, at 20~cm and 90~cm, into three non-thermal
sources and regard them as likely AGN, although they note the
nearby ``unrelated'' IRAS source just 1\arcmin\ to the north (IRAS~18352$-$0655);
the IRAS source represents the cumulative flux from the RSGs. Fig.~\ref{fig:contour_20cm}
shows flux from the radio objects in the field at 20~cm from the Galactic Plane
Survey \citep{whi05,hel05}\footnote{The data for this figure are from 
MAGPIS: The Multi-Array Galactic Plane Imaging Survey (http://third.ucllnl.org/gps/index.html).}. 
We discuss these three sources in detail below. In general, the sources are non-thermal
in nature, supporting our claim that they are related to supernovae activity from the
cluster. They are also spatially coincident with the cluster, and similar radio sources
are not seen within a one square degree field in the MAGPIS data, again supporting a 
connection between the radio sources and the cluster. 

\clearpage
\citet{zoo90} identify GPSR~25.267$-$0.161, near star \#6, in 21.4~cm observations using the VLA in the
B configuration, i.e.\ positional accuracy of $\sim3\arcsec$. This is the
same object as GPSR5~25.266$-$0.161, identified by \citet{bec94} at 6~cm. 
The source has a spectral index of 
$\alpha=$log(254~{\it mJy}/551~{\it mJy})/log(5~{\it GHz}/1.4~{\it GHz})=$-$0.60 \citep{hel89b,bec94}. 

\citet{bec94} associate two separate sources with GPSR5~25.252$-$0.139, having
spectral indices of 0.14 and $<-$1.6, and separated by 7\arcsec\ in the north-south 
direction. These sources appear to be at the center of emission that extends 1\arcmin\ to the north and
south at 20~cm. It is intriguing to note that its size, $\sim1\arcmin\times2\arcmin$, is
roughly the same the Crab nebula would have at 5,800~pc. The size of the 
Crab is $\sim4\arcmin\times6\arcmin$,
and it is at a distance of 1930~pc. At 5800~pc, its apparent size would be $\sim1.3\arcmin\times2.0\arcmin$.
The extended emission does not appear at 6~cm in the Galactic Plane Survey at a level above 
1~mJy per beam, or 28~$\mu${\it Jy~arcsec$^{-2}$}. This non-detection suggests a spectral
index of 
$\alpha<$log(0.028~{\it mJy}/149~{\it mJy})/log(5~{\it GHz}/1.4~{\it GHz})=$-$1.32, i.e.\
the source of the extended emission is non-thermal. We suggest that this source may be a
supernovae remnant, similar to the Crab. 

\citet{bec94} also identify GPSR5~25.237$-$0.150 about 1\arcmin\ to the south of star \#4 and
GPSR~25.267$-$0.161. This source has a spectral index of \\
$log(111~{\it mJy}/238~{\it mJy})/log(5~{\it GHz}/1.4~{\it GHz})=-$0.60.

The Culgoora far-infrared/sub-millimeter circular array identified
a prominent source in the field, Cul1835$-$069 \citep{sle95}, with an effective beam
diameter of several arcminutes. The spectral index at these wavelengths is $-$0.70. Given the
similarity of this spectral index to that at cm wavelengths, we assume that the radio and submillimeter
objects are, in fact, the same, and different from the flux from the RSGs ($\alpha$=2 for the Rayleigh-Jeans tail
of a blackbody). 
\citet{vol89} identify NSV11126 as having an unusual IRAS LRS spectrum at mid-infrared wavelengths.
This source likely represents light from the RSGs, as do IRAS~18352$-$0655 and RAFGL~5268S. 
As already discussed, the cluster is identified in {\it MSX} images at mid-infrared wavelengths 
(see Fig.~\ref{fig:msx}).

\subsection{End States and Initial Stellar Mass}

A number of studies predict that the most massive stars will largely evaporate in
their lifetimes, leaving relatively low mass objects before they explode as supernovae. 
For example, \citet{heg03} predict that stars with solar metallicities and \Minit$>$50~\Msun\ will collapse as neutron
stars, whereas those with 25~\Msun$>$\Minit$>$50~\Msun\ collapse as black holes. Lower mass
stars, 9~\Msun$>$\Minit$>$25~\Msun, will collapse as neutron stars. 
These models are difficult to test. A successful test requires that an object have a well determined initial mass and
that the nature of the end state is well known. Of course, once the object has reached the end state, it
is difficult to independently infer the initial mass of that object. In a few cases, a supernova progenitor has been 
observed, but even in these cases, there is usually a very poor correlation between progenitor and
initial mass \citep{sma04,van05}. These progenitors have always been either RSGs or BSGs, and it is
difficult to uniquely determine initial masses for such stars. 

One promising approach to relate end state to initial mass is to find a recent supernovae remnant and
relate the object to the present day upper mass cutoff in the coeval cluster in which it was formed. 
This appraoch has been successfully used to relate magnetars to particularly massive 
progenitors (\Minit$>$50~\Msun) \citep{fig05b,mun05}, and an alternate method confirms these
results \citep{gae05}. In this regard, the RSG cluster might be 
an ideal test bed for such a connection. Given
that the most massive stars are RSGs, we expect there to be a very narrow range of
masses defining the most massive stars in the cluster, although the precise value of that range
will scale with age (Fig.~\ref{fig:model}). In any case, it is certainly likely to be somewhere
between 12 and 25~\Msun, with a bias for a younger age and masses at the high end of the range. 

We plan to obtain deeper observations in order to provide a tighter constraint on the masses of 
the stars (and, in turn, the cluster age). We also plan to determine the nature of the most likely
compact object that might have been produced in the cluster. 

\begin{figure}
\epsscale{1.15}
\plotone{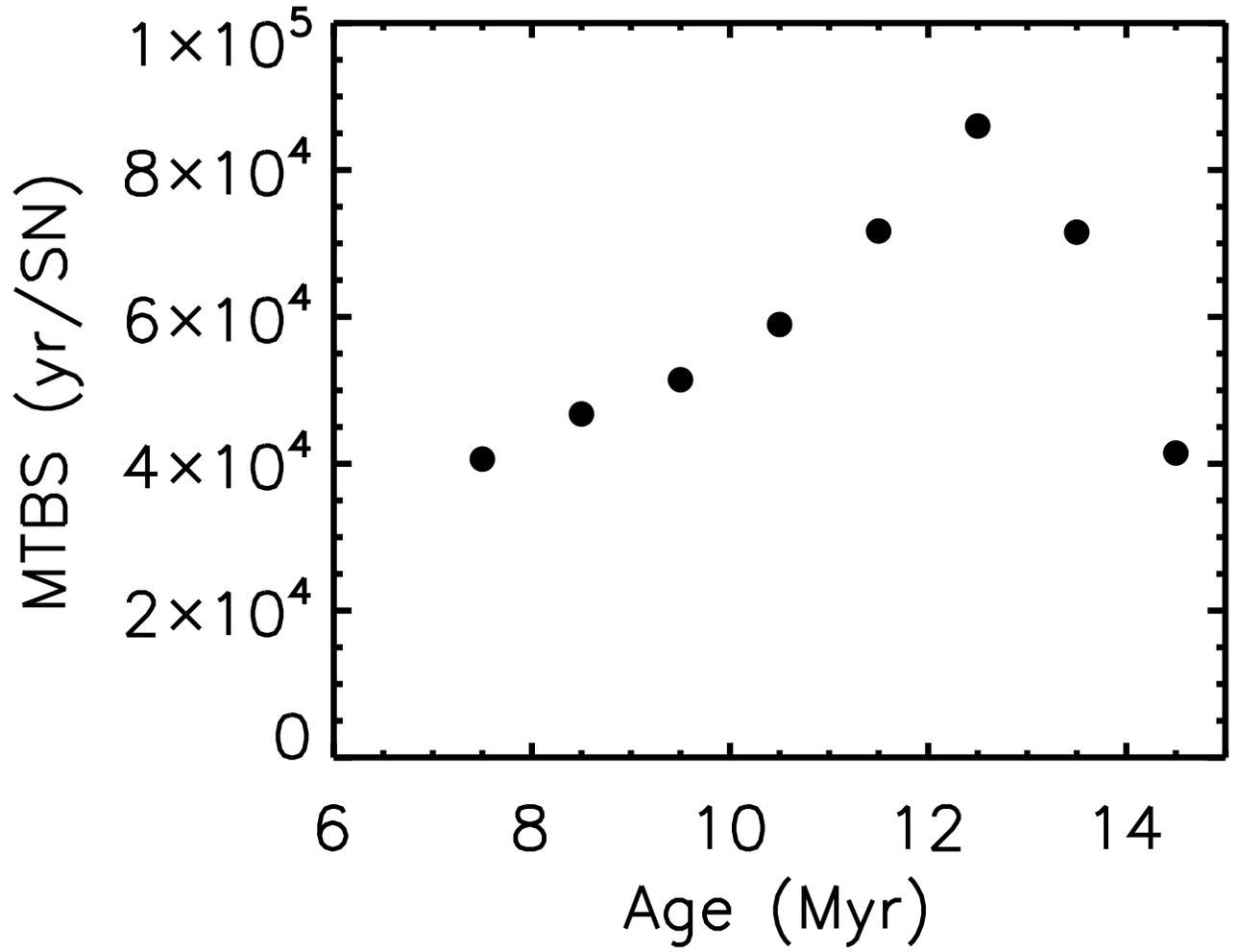}
\caption
{\label{fig:sim} Mean time between supernovae for a cluster with 14 RSGs versus age.}
\end{figure}

We can estimate the supernova rate for this cluster as a function of assumed age. As with our
previous simulations, we assume 14 RSGs and the Geneva models with solar metallicity and
the canonical mass-loss rates. Fig.~\ref{fig:sim} shows the average time between supernovae for
clusters that have enough mass (Fig.~\ref{fig:mass_monte}) to produce 14 RSGs at a given age. 
We find that there should be a supernova in the cluster every $\sim$40,000 to 80,000~yr, on average. This
timescale is consistent with the presence of putative post-supernovae phenomena, i.e.\ x-ray and
$\gamma$-ray sources, as found near the cluster. 
\clearpage
\clearpage

\subsection{Comparison to Other Clusters with RSGs}
The cluster mass is extraordinary, perhaps trailing only Westerlund~1, as the most massive young cluster
in the Galaxy \citep{cla05}. One difficulty in making cluster mass comparisons is that all of
the mass estimates are made by extrapolating an IMF
down to masses below the observational limit. Only in the Arches cluster, near the Galactic
center, do we directly count the majority of the stars in the cluster. In that case, the
slope of the initial mass function, for stars presently in the cluster, deviates significantly from
the Salpeter value \citep{fig99a,sto02}. Thus, the initial cluster mass for the Arches cluster
is inferred to be $\sim$10,000~\Msun, even though it presently has an estimated 160 O-stars \citep{fig02}. 
However, it is possible that the true initial mass was
much higher and that the present-day cluster has been heavily evacuated of low-mass stars via
N-body interactions \citep{kim00,por02}. If we simply integrate a Salpeter IMF from the high
mass stars down to 0.5~\Msun, we find a mass of 50,000~\Msun. Using this technique, we would
find similar values for the Quintuplet cluster and the cluster in the central parsec of
the Galaxy \citep{fig99b}. For Westerlund~1, we would find a mass of $\sim$10$^5$~\Msun\ \citep{cla05}.
Given the uncertainties of all the estimates involved, we claim that the cluster of RSGs
is in a class of massive young stellar clusters containing the aforementioned clusters. 
Until these clusters are observed down to $\sim$1~\Msun, and the effects of dynamical evolution better understood, 
it will be difficult to know their true initial mass-rank ordering. 

In any case, this group of clusters substantially overlaps with the low mass end of the super star clusters
often found in interacting galaxies \citep{whi99}, and the low mass end of the Galactic globular
cluster population \citep{man91}. This indicates that the cluster formation mechanisms presently found in 
the Galaxy might be similar to those in interacting systems or in the young Galaxy, albeit at
a much subdued level.  

\subsection{Comparison to NGC~7419}
One can compare this cluster to NGC~7419, previously thought to be the richest cluster of red
supergiants in the Galaxy, with five. 
\citet{car03} note 
that the BSG to RSG ratio in this cluster is lower
than that for any known cluster in the Galaxy, at
least as low as 1:5, and may be due to the often mentioned effects of the observed 
rapid stellar rotation in the stars \citep{mae01}. Model isochrones successfully 
match the observations for an age of 15~\Myr. This cluster has an estimated
metallicity near the solar value. 

\begin{figure}
\epsscale{1.1}
\plotone{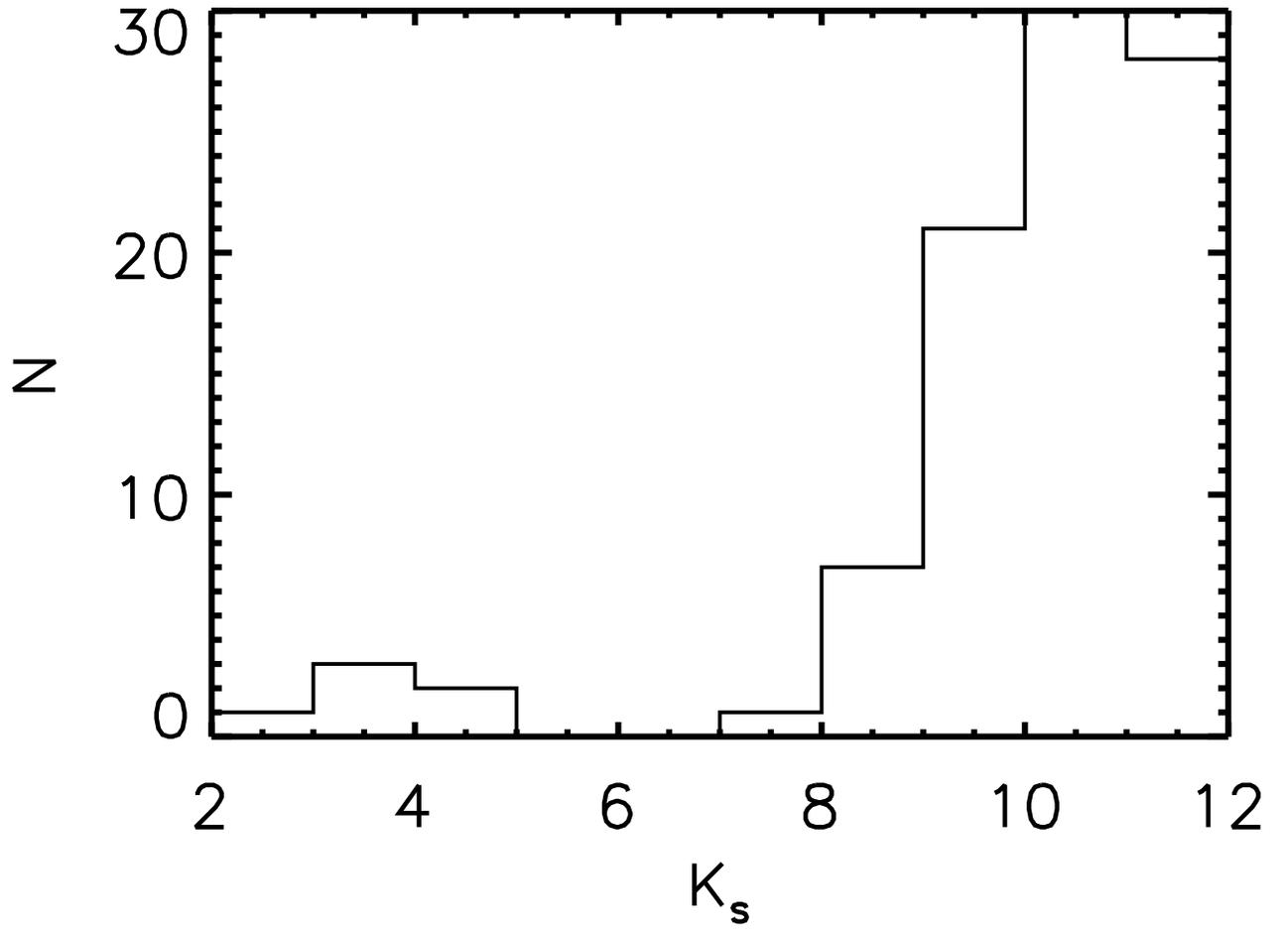}
\caption
{\label{fig:hist7419} {\it 2MASS} luminosity function of stars in the field containing NGC~7419. The sample has been culled
of stars with reported errors greater than 0.1 magnitudes in {\it K}$_s$.}
\end{figure}

\begin{figure*}
\plottwo{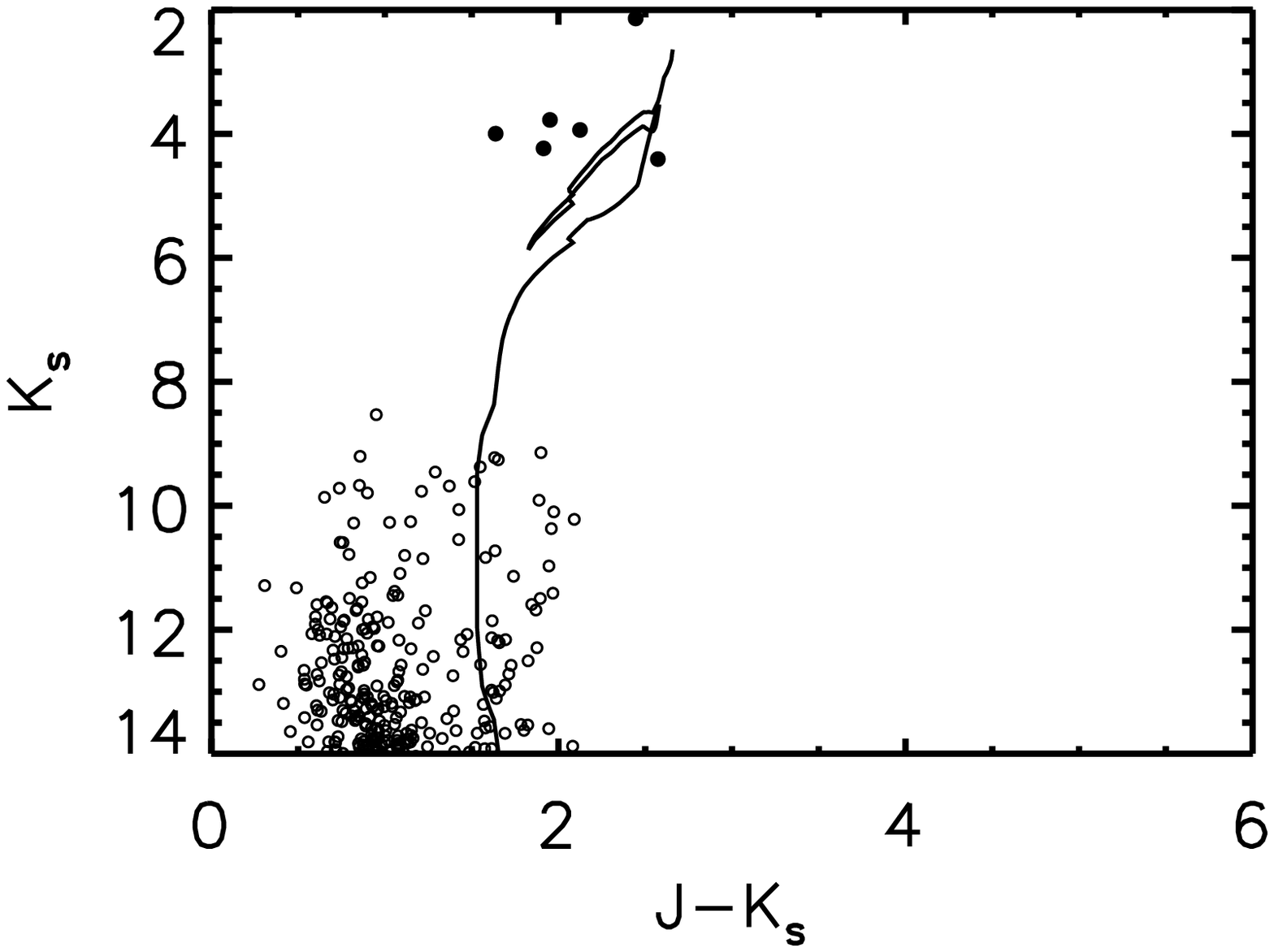}{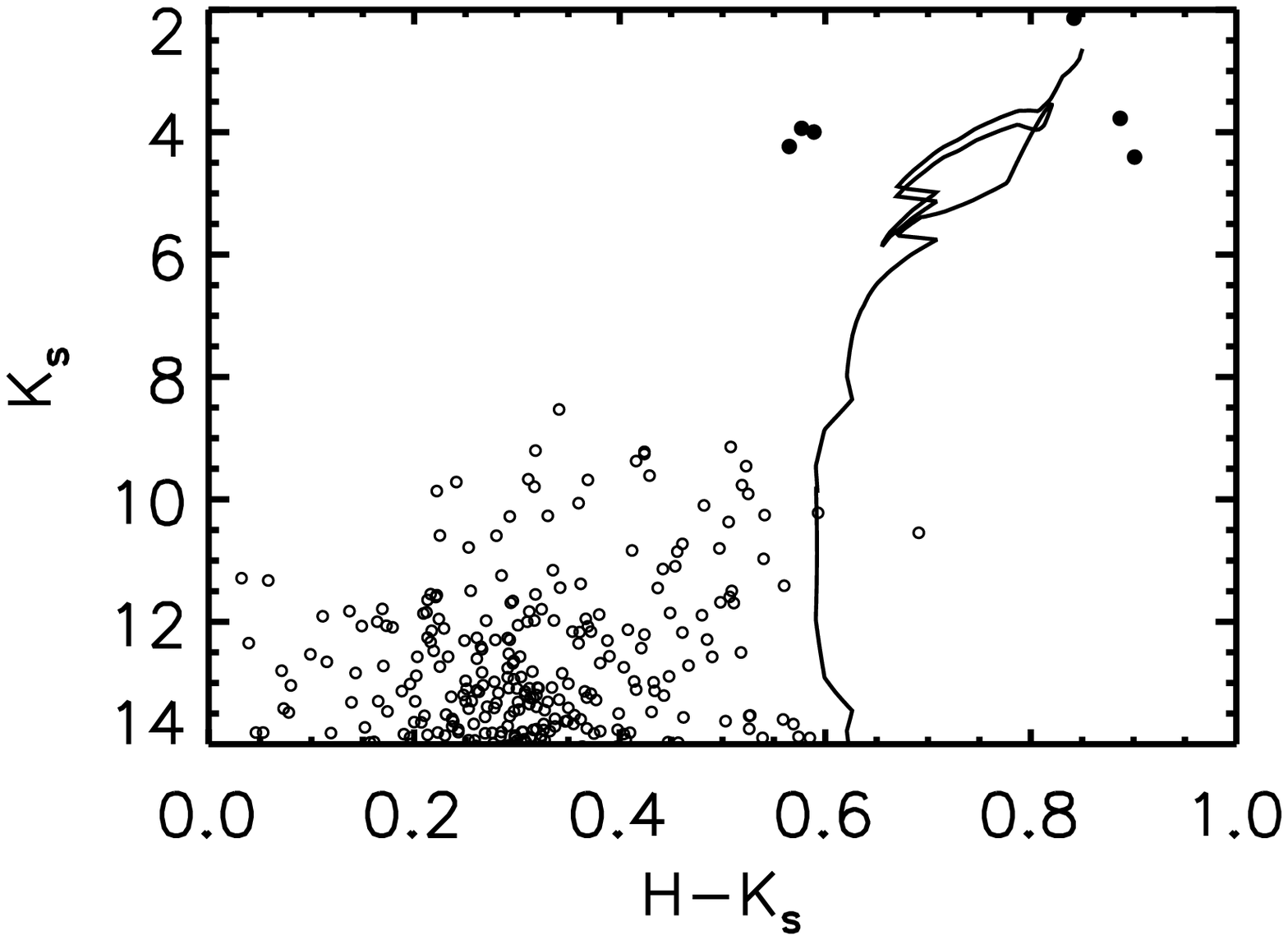}
\caption
{\label{fig:cmd7419} {\it 2MASS} color-magnitude diagram of stars in the field containing NGC~7419.
The RSGs are represented by filled circles, and the fainter field stars 
are designated by open circles. A 15~\Myr\ isochrone is overplotted, assuming
the Geneva models with solar metallicity, a distance of 2.3~\kpc, A$_{K_s}$=0.7, and
the redenning law of \citet{rie89}. The sample has been culled
of stars with reported errors greater than 0.1 magnitudes in {\it K}$_s$.}
\end{figure*}

\begin{figure*}
\plottwo{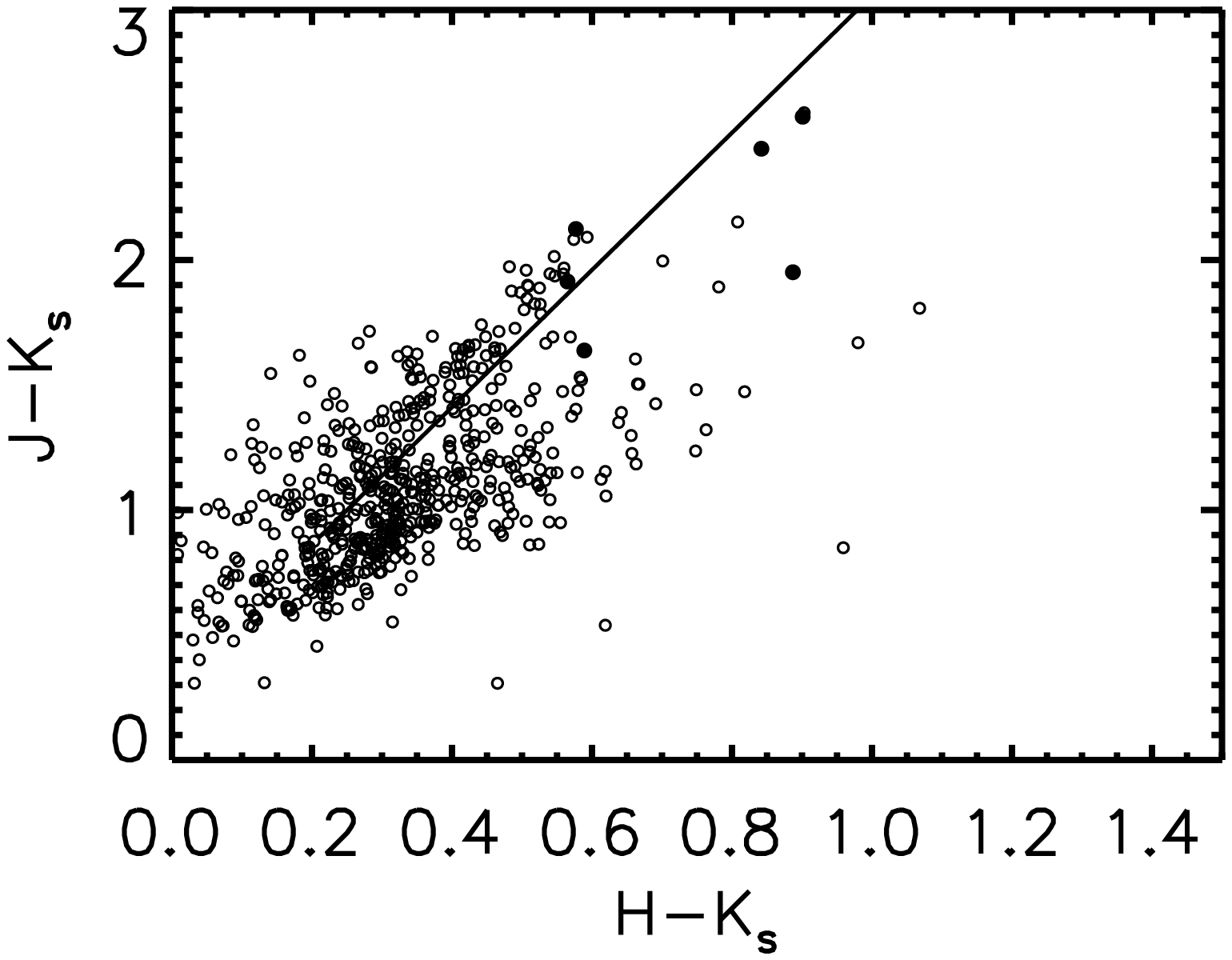}{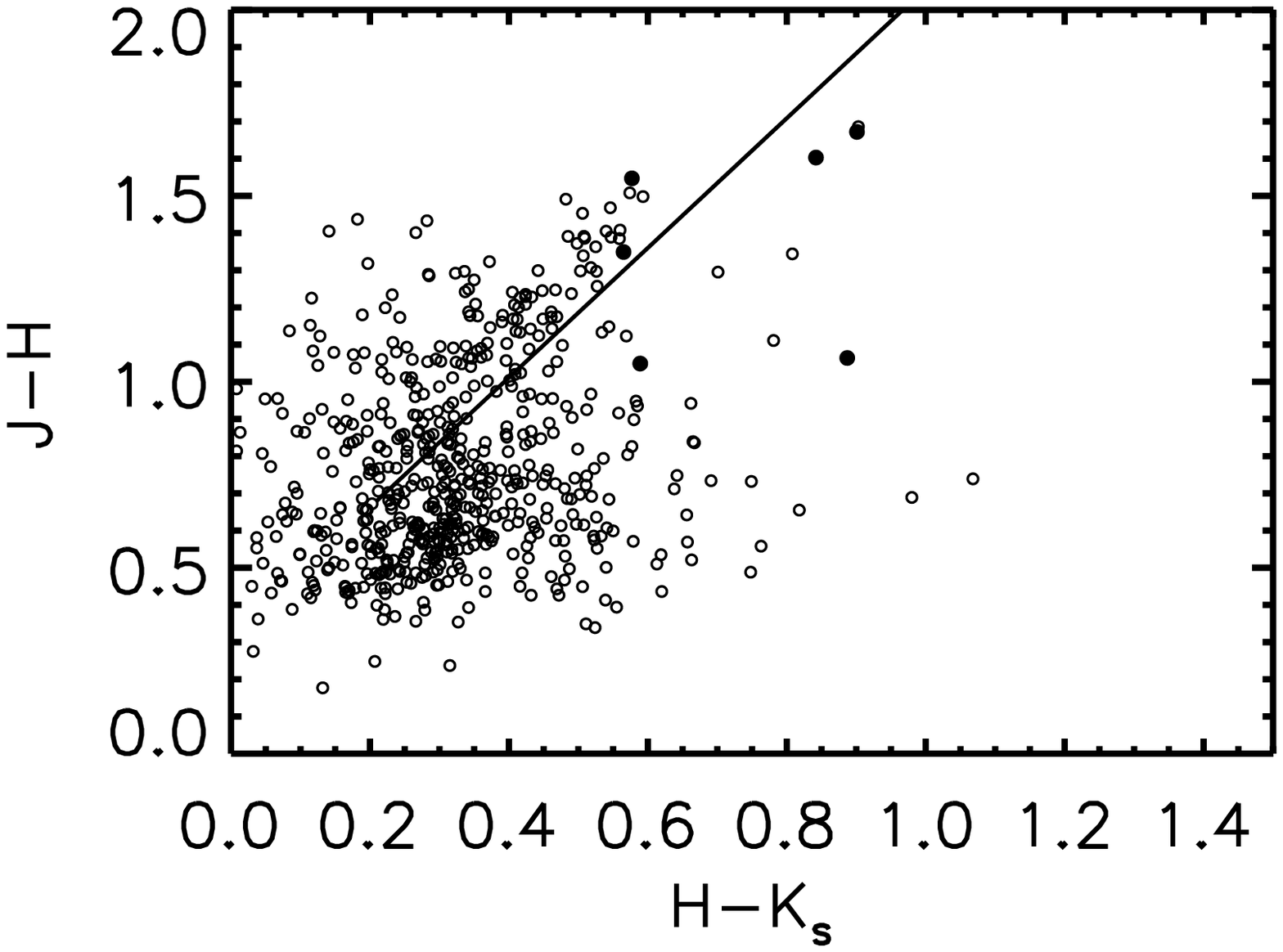}
\caption
{\label{fig:ccJK7419} {\it 2MASS} color-color diagrams of stars in the field containing NGC~7419. The 
RSGs are represented by filled circles, and the fainter field stars 
are designated by open circles. A redenning vector is plotted from the expected 
intrinsic colors of RSGs, assuming the extinction law of \citet{rie89}. The sample has been culled
of stars with reported errors greater than 0.1 magnitudes in {\it K}$_s$.}
\end{figure*}

Figs.~\ref{fig:hist7419}, \ref{fig:cmd7419}, and \ref{fig:ccJK7419}, show
{\it 2MASS} data for NGC~7419 that are similar to those seen for the 
RSG cluster discussed in this paper. That is, there
appears to be a large gap in K$_s$ between the RSGs and the bulk of the fainter
stars in the cluster. 

\begin{figure}
\epsscale{1.1}
\plotone{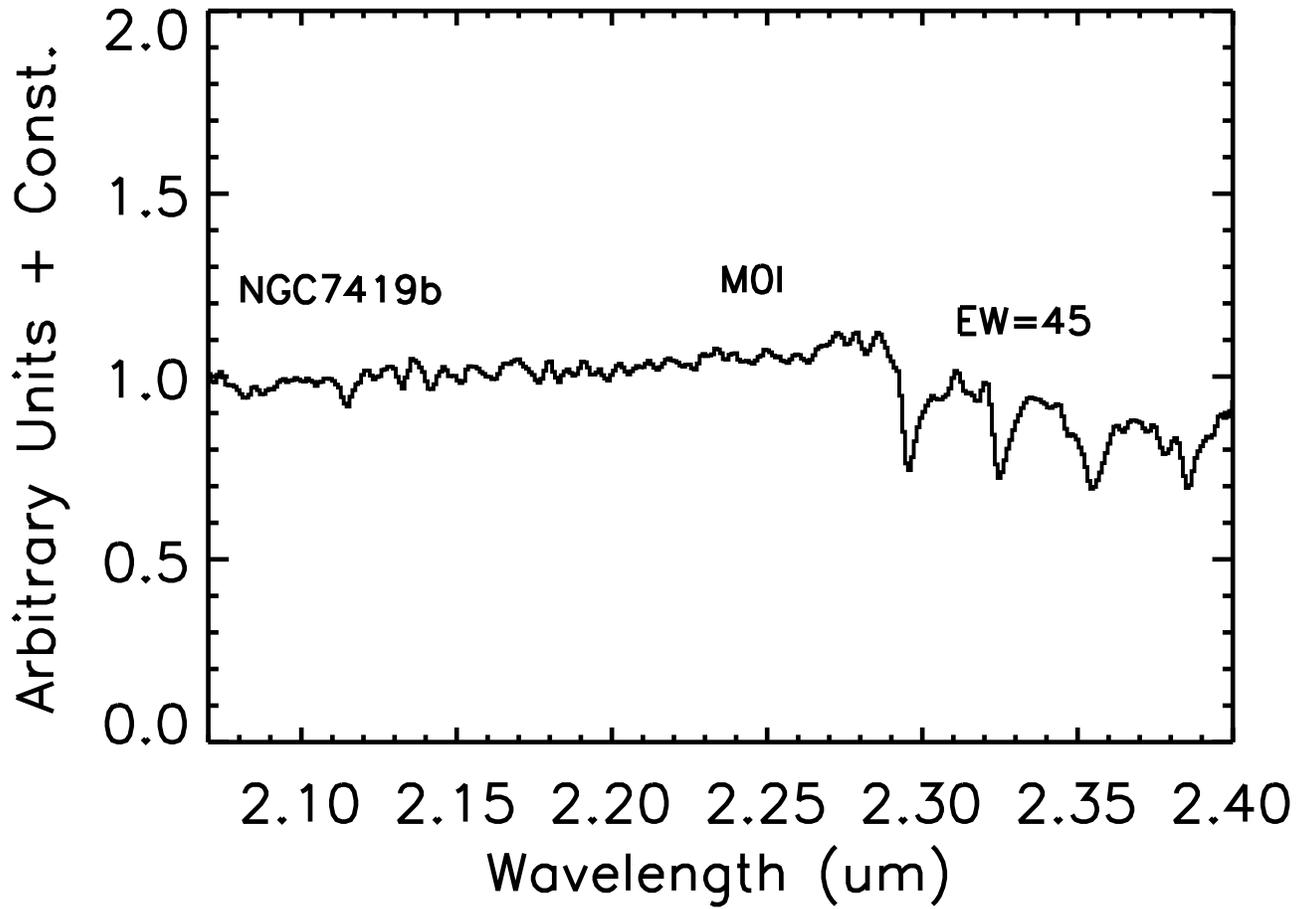}
\caption
{\label{fig:NGC7419spec} {\it IRMOS} spectrum of NGC~7419b. \citet{bla55} give a spectral type of M2~I using optical data, whereas
we estimate M0~I with our method.}
\end{figure}

We obtained a spectrum of one of the RSG stars (Fig.~\ref{fig:NGC7419spec})
which indicates a spectral type of M0~I, compared to M2~I, given by \citet{bla55}. This validates
our method to an accuracy within a few subtypes. 

Assuming a distance of 2~\kpc\ to NGC~7419, the average of the values in \citet{car03} and \citet{bea94}, and
\AK=0.67$\times$0.112 \citep{bea94}, we find M$_{\rm K}=-7.6$ for the four fainter RSGs, and
M$_{\rm K}=-9.5$ for the brightest member. The extraordinary infrared brightness of the brightest member
can be attributed to its very red spectral energy distribution (M7.5~I); note that this star also
produces an OH maser, just as we find for one of the stars in our cluster. 
In comparison, the stars in our cluster have $M_{\rm K}\sim-$10.4~to~$-$11.6 (see Table~1). 
\citet{bea94} note that the RSGs in NGC~7419 are faint compared to those found elsewhere in the
Galaxy that typically have $M_{\rm K}\sim-10.5$. We interpret the data to indicate that our
cluster has relatively massive stars, and is thus younger than 15~\Myr, the age of NGC~7419 \citep{car03}.
This is consistent with Fig.~\ref{fig:model} which shows that $M_{\rm K}$ 
progresses from $\sim-11.2~to~-10.2$ for the brightest RSG in a cluster with an 
$\tau_{\rm age}\sim$8~to~18~\Myr, respectively.  

\clearpage

\clearpage
\subsection{Comparison to h and $\chi$ Persei and the Per OB1 Association}

h and $\chi$~Per, and the surrounding Per~OB1 association, 
collectively contain $\sim$20 RSGs. While containing more RSGs than in 
the cluster we describe, the Perseus
RSGs were likely formed in several episodes spanning on order of 10~\Myr\ \citep{sle02}. Because of
this, it is difficult to directly analyze the properties of the full Perseus RSG
population without directly knowing which members belong to the various star formation episodes. 
Individually, h and $\chi$~Per are each $\sim$10~\Myr\ old and contain $\sim$5000~\Msun\ in stellar mass \citep{bra05}, considerably less
than estimated for our cluster. We presume this is consistent with the relatively low number of
bona fide RSGs (only one) that can directly be attributed to either of the two double clusters,
whereas our cluster has 14.

\section{Conclusions}
We have identified one of the most massive young stellar clusters in the Galaxy, containing
14 RSGs, more than in any cluster in the LMC, SMC, or the Galaxy. From an 
analysis of the estimated luminosities of the RSGs, we estimate an age of $\sim$10~\Myr.
We further note the rich collection of radio, x-ray, and $\gamma$-ray, objects coincident
with the location of the cluster that can be understood as a natural consequence of massive
stars progressing to supernovae, as predicted by stellar evolution models. The RSGs
in this cluster are likely direct progenitors of supernovae and provide a particularly ripe
sample for studies relating compact objects to initial stellar masses. 
We note that one particular object, GPSR5~25.252$-$0.139, is a likely SNR, with properties similar
to the Crab nebula. 
In order to test stellar evolution models, we plan to obtain deeper 
observations of the RSG cluster to estimate
the IMF slope, and BSG to RSG ratio. We also plan to obtain spectroscopic observations in order
to determine if the rotation rates of the RSGs are consistent with theoretical 
expectations for such a rich cluster of these relatively rare stars. 

\acknowledgements
We appreciate useful discussions with Mike Muno, Andy Fruchter, Brad Whitmore, Claus Leitherer, Nolan Walborn, 
Rick White, Roberta Humphreys, Phil Massey, Norbert Langer, Tony Moffat, and Andre Maeder. We 
thank the Spitzer/GLIMPSE team, especially Ed Churchwell, Marilyn Meade, Brian Babler, Remy Indebetouw, and Barbara Whitney, 
for making images available before public release.
We also thank the referee for useful comments, especially regarding the distance to W42. 
The material in this paper is based upon work supported by NASA
under award No.\ NNG05-GC37G, through the {\it Long Term Space Astrophysics} program.
F.N. acknowledges PNAYA2003-02785-E and AYA2004-08271-C02-02 grants and 
 the Ramon y Cajal program. {\it IRMOS} is supported by {\it NASA/JWST, NASA/GSFC, STScI DDRF}, and {\it KPNO}.
This research has
made use of data obtained from the High Energy Astrophysics Science Archive Research
Center ({\it HEASARC}), provided by NASA's {\it Goddard Space Flight Center}.
This work made use of data from, 
{\it MAGPIS: The Multi-Array Galactic Plane Imaging Survey} \citep{whi05,hel05},
and the {\it MAGPIS} web site: http://third.ucllnl.org/gps/index.html.

\end{document}